\documentclass[usenatbib, usedcolumn]{mn2e}
\usepackage{lscape}

\title[The mass function of NGC 2547]{Low mass stars, brown dwarf
  candidates and the mass function of the young open cluster NGC 2547}
\author[Jeffries et al.]
       {R.D. Jeffries$^1$, Tim Naylor$^2$, C.R. Devey$^1$,
     and E.J. Totten$^1$
\\
$^1$Astrophysics Group, School of Chemistry and Physics, 
Keele University, Staffordshire, ST5 5BG \\
$^2$School of Physics, University of Exeter, Stocker Road, Exeter EX4 4QL
}
\date{}

\pagerange{\pageref{firstpage}--\pageref{lastpage}}
\pubyear{}

\begin{document}
\newcommand{\rmsub}[2]{#1_{\rm #2}} 
\newcommand{\alfven}{Alfv\'en}
%
%
\newcommand{\aat}{\mbox{\em AAT}}
\newcommand{\eso}{\mbox{\em ESO}}
\newcommand{\iue}{\mbox{\em IUE}}
\newcommand{\exosat}{\mbox{\em EXOSAT}}
\newcommand{\einstein}{\mbox{\em Einstein}}
\newcommand{\ginga}{\mbox{\em GINGA}}
\newcommand{\rosat}{\mbox{\em ROSAT}}
\newcommand{\caspec}{\mbox{\em CASPEC}}
\newcommand{\ucles}{\mbox{\em UCLES}}
\newcommand{\starlink}{\mbox{\em Starlink}}
\newcommand{\etal}{\mbox{\em et\ al.\ }}
\newcommand{\dex}[1]{\hbox{$\times\hbox{10}^{#1}$}}
\newcommand{\eex}[1]{\hbox{$\hbox{10}^{#1}$}}
\newcommand{\gpar}{\mbox{$g_{\parallel}$}}
\newcommand{\vsi}{\mbox{$v_e\,\sin\,i$}}
\newcommand{\vsini}{\mbox{$v_e\,\sin\,i$}}
\newcommand{\pattcit}[1]{\hbox{$^{(#1)}$}}
\newcommand{\pattcite}[1]{\hbox{$^{(#1)}$}}
%
%
\newcommand{\ha}{H$\alpha$}
\newcommand{\hb}{H$\beta$}
\newcommand{\hgam}{H$\gamma$}
\newcommand{\hdel}{H$\delta$}
\newcommand{\heps}{H$\epsilon$}
\newcommand{\lya}{\hbox{$\hbox{Ly}\alpha$}}
\newcommand{\naid}{\mbox{Na{\footnotesize I} {\sl D}}}
\newcommand{\caii}{Ca\,{\footnotesize II}}
\newcommand{\caiih}{Ca\,{\footnotesize II}~H}
\newcommand{\caiik}{Ca\,{\footnotesize II}~K}
\newcommand{\caiihk}{Ca\,{\footnotesize II}~H \&~K}
\newcommand{\mgii}{Mg\,{\footnotesize II}}
\newcommand{\mgiih}{\mbox{Mg{\footnotesize II} {\sl h}}}
\newcommand{\mgiik}{\mbox{Mg{\footnotesize II} {\sl k}}}
\newcommand{\mgiihk}{\mbox{Mg{\footnotesize II} {\sl h} \&\ {\sl k}}}
\newcommand{\lii}{Li\,{\footnotesize I}}
\newcommand{\fei}{Fe\,{\footnotesize I}}
\newcommand{\baii}{Ba\,{\footnotesize II}}
\newcommand{\ki}{K\,{\footnotesize I}}
\newcommand{\cai}{Ca\,{\footnotesize I}}
%
%
\newcommand{\ang}{\,\mbox{\AA}}
\newcommand{\Ang}{\ang}
\newcommand{\angstrom}{\ang}
\newcommand{\angstroms}{\ang}
\newcommand{\Angstrom}{\ang}
\newcommand{\Angstroms}{\ang}
\newcommand{\km}{\,km}
\newcommand{\Mpc}{\,\mbox{Mpc}}
\newcommand{\kpc}{\,\mbox{kpc}}
\newcommand{\kms}{\,km\,s$^{-1}$}
\newcommand{\ergs}{\,\mbox{$\mbox{erg}\,\mbox{s}^{-1}$}}
\newcommand{\ergsqcmsecang}{\,erg\,cm$^{-2}$\,s$^{-1}$\,\AA$^{-1}$}
\newcommand{\ergsqcm}{\,erg\,cm$^{-2}$}
\newcommand{\ergsqcmsec}{\,erg\,cm$^{-2}$\,s$^{-1}$}
\newcommand{\sqcm}{\,\mbox{$\mbox{cm}^{2}$}}
\newcommand{\cucm}{\,\mbox{$\mbox{cm}^{3}$}}
\newcommand{\persqcm}{\,\mbox{$\mbox{cm}^{-2}$}}
\newcommand{\gpersqcm}{\,\mbox{g}\persqcm}
\newcommand{\percc}{\,\mbox{$\mbox{cm}^{-3}$}}
\newcommand{\kev}{\,\mbox{keV}}
\newcommand{\kelvin}{\,K}
\newcommand{\kgmcube}{\,\mbox{$\mbox{kg}\,\mbox{m}^{-3}$}}
\newcommand{\dynsqcm}{\,\mbox{$\mbox{dyn}\,\mbox{cm}^{-2}$}}
\newcommand{\degrees}{\mbox{$^\circ$}}
\newcommand{\rstar}{\,\mbox{$\mbox{R}_*$}}
\newcommand{\mstar}{\,\mbox{$\mbox{M}_*$}}
\newcommand{\lstar}{\,\mbox{$\mbox{L}_*$}}
\newcommand{\vstar}{\,\mbox{$\mbox{V}_*$}}
\newcommand{\msun}{\,\mbox{$\mbox{M}_{\odot}$}}
\newcommand{\rsun}{\,\mbox{$\mbox{R}_{\odot}$}}
\newcommand{\lsun}{\,\mbox{$\mbox{L}_{\odot}$}}
\newcommand{\lx}{\,\mbox{$L_{\rm x}$}}
%
%
\newcommand{\reference}[5]{\noindent #1, #2. {\sl #3\/}, {\bf #4,} \,\mbox{#5}} 
\newcommand{\refnum}[4]{\noindent #1, #2. {\sl #3\/}, {\bf #4}}
\newcommand{\refbook}[3]{\noindent #1, #2. {\sl #3\/}}
\newcommand{\refpress}[3]{\noindent #1, #2. {\sl #3\/}, in press}
\newcommand{\refsub}[3]{\noindent #1, #2. {\sl #3\/}, submitted}
\newcommand{\refprep}[2]{\noindent #1, #2. In preparation}
%
%
\newcommand{\aanda}  {Astr.\ Astro\-phys.\nolinebreak\ }
\newcommand{\aasupp} {Astr.\ Astro\-phys.\nolinebreak\ Suppl.\nolinebreak\ }
\newcommand{\aj}     {Astron.\nolinebreak\ J.\nolinebreak\ }
\newcommand{\annrev} {Ann.\ Rev.\ Astr.\ Astro\-phys.\nolinebreak\ }
\newcommand{\acta}   {Acta Astron.\nolinebreak\ }
\newcommand{\apj}    {Astro\-phys.\nolinebreak\ J.\nolinebreak\ }
\newcommand{\apjs}   {Astro\-phys.\nolinebreak\ J.\ Suppl.\nolinebreak\ }
\newcommand{\apjsupp}{\apjs}
\newcommand{\aplett} {Astro\-phys.\nolinebreak\ Lett.\nolinebreak\ }
\newcommand{\gafd}   {Geo\-phys.~Astro\-phys.\ Fluid Dyn.\nolinebreak\ }
\newcommand{\ibvs}   {Inf.\ Bull.\ var.\ Stars\nolinebreak\ }
\newcommand{\jgr}    {J.\ Geo\-phys.~Res.\nolinebreak\ } 
\newcommand{\jpp}    {J.\ Plasma Phys.\nolinebreak\ }
\newcommand{\mn}     {Mon.\ Not.\ R.\ astr.\nolinebreak\ Soc.\nolinebreak\ }
\newcommand{\pf}     {Phys.\nolinebreak\ Fluids\nolinebreak\ }
\newcommand{\pasp}   {Publ.\ astr.\ Soc.\ Pacif.\nolinebreak\ }
\newcommand{\sovast} {Soviet astr.\nolinebreak\ }
\newcommand{\procasa}{Proc.\ Astr.\ Soc.\ Australia\nolinebreak\ }
\newcommand{\solp}   {Solar Phys.\nolinebreak\ }
%
%
\newcommand{\deriv}[2]{\mbox{${{\displaystyle d#1}\over
                       {\displaystyle d#2}}$}} 
\newcommand{\sderiv}[2]{\mbox{${{\displaystyle d^2#1}\over
                       {\displaystyle d#2^2}}$}} 
\newcommand{\pderiv}[2]{\mbox{${{\displaystyle\partial#1}\over
                       {\displaystyle\partial#2}}$}} 
\newcommand{\spderiv}[2]{\mbox{${{\displaystyle\partial^2#1}\over
                        {\displaystyle\partial#2^2}}$}} 
\newcommand{\half}{\mbox{$\frac{1}{2}$}}
\newcommand{\twiddles}{\mbox{$\sim $}}
\newcommand{\varomega}{\varpi}
\newcommand{\twid}{\mbox{$\sim $}}
\newcommand{\bvec}[1]{\mbox{\boldmath ${#1}$}}
\newcommand{\be}{\begin{equation}}
\newcommand{\ee}{\end{equation}}
\newcommand{\bd}{\begin{displaymath}}
\newcommand{\ed}{\end{displaymath}}

\maketitle

\label{firstpage}

\begin{abstract}
We present a catalogue of $R_{\rm c}I_{\rm c}Z$ photometry over an area
of 0.855 square degrees, centred on the young open cluster NGC 2547.
The survey is substantially complete to limits of $R_{c}=21.5$, $I_{\rm
c}=19.5$, $Z=19.5$. We use the catalogue to define a sample of NGC 2547
candidates with model-dependent masses of about 0.05-1.0\,$M_{\odot}$.
After correcting for incompleteness and estimating contamination by
foreground field dwarfs, we investigate the mass function of the
cluster, its binary content, and search for evidence of mass
segregation among the lower mass stars.  There is ample evidence for
mass segregation between high ($>3\,M_{\odot}$) and lower mass stars,
but over the range $0.1<M<0.7\,M_{\odot}$, the data are consistent with
no further mass segregation. By fitting King profiles we conclude that
at least 60 percent of the low-mass stellar population are contained
within our survey. The cluster mass function is remarkably similar to
the Pleiades for $0.075<M<0.7\,M_{\odot}$.  Because of its age ($\simeq
30$\,Myr), we demonstrate that this mass function is robust to a number
of systematic uncertainties likely to affect older and younger clusters
and is therefore one of the best available estimates for the initial
mass function in young disc populations.  For $0.05<M<0.075\,M_{\odot}$
there is some evidence for a deficit of brown dwarfs in NGC 2547
compared with other clusters. This deficit may extend to lower masses
or may only be a dip, perhaps caused by an imperfect understanding of the
mass-magnitude relationship at temperatures of around 2800\,K.
Incompleteness in both our survey, and the luminosity functions from
which we estimate contamination by foreground objects, leave this
question open. The binary fraction for systems with mass ratios greater
than about 0.5 is 20-35 per cent for M-dwarfs in NGC 2547, quite
consistent with that found in the field and other young clusters.  The
full photometric catalogue and our lists of candidate cluster members
are made available in electronic format.
\end{abstract}


\section{INTRODUCTION}
\label{intro}

The open cluster NGC 2547 (= C0809-491) is important for efforts to
understand the formation and evolution of low-mass, pre-main-sequence
(PMS) stars.  Its isochronal age is about 30\,Myr, it is relatively
close ($\simeq$ 400--450\,pc), has low reddening ($E(B-V)=0.06$) and a
rich population of identified early and late-type members (see
Clari\'{a} 1982; Jeffries, Totten \& James 2000; Naylor et al. 2002 --
hereafter N02). Being close and quite compact it serves as an ideal
laboratory to investigate evolutionary tracks and isochrones, rotation
rates, magnetic activity and light element abundances of low-mass PMS
stars as they approach the ZAMS. In particular, NGC 2547 occupies an
important and comparatively little studied niche of parameter space
with an age between those of regions of recent star formation, and
clusters like the Pleiades, where low-mass stars have already reached
the zero-age main-sequence (ZAMS).

Jeffries \& Tolley (1998) first identified the low-mass PMS population
of NGC 2547 as counterparts to {\em ROSAT} X-ray sources. A number of
these counterparts were positively identified as F to K-type cluster
members by Jeffries et al. (2000) using radial velocities. A $BVI_{\rm
c}$ photometric study of NGC 2547 was presented by N02. 
Their survey covered an approximately square area of 0.32
square degrees and identified several hundred candidate cluster members
with masses from 6\,$M_{\odot}$ down to about 0.25$M_{\odot}$
at $V\simeq20.5$. N02 concluded that the mass function
(MF) was similar in shape to that of the Pleiades and of
field stars over this mass range, but that the total mass of NGC 2547
could be a few times less than the Pleiades. Littlefair et
al. (2003) have used the same list of candidate members to show that
stars with mass $>3\,M_{\odot}$ in NGC 2547 are much more
centrally concentrated than lower mass objects. Because NGC 2547 is
no more than 10 dynamical crossing times old, it seems likely that most of
this segregation must be primordial (e.g. Bonnell et al. 2001),
rather than due to dynamical evolution.

The N02 survey did not cover a sufficiently large area or go to
sufficient depths to discover whether the spatial distribution of
low-mass ($<1\,M_{\odot}$) cluster members was independent of mass.
The possibility of differential mass segregation between lower mass
stars is important. Some recent theories concerning the formation of
very low-mass stars and brown dwarfs suggest they are ejected as
low-mass fragments from protostellar multiple systems before they have
a chance to accrete significant material (e.g. Reipurth \& Clarke
2001). Such objects may have a greater velocity dispersion than their
higher mass siblings and hence be less spatially concentrated in a
cluster (Sterzik \& Durisen 2003).  This in turn would mean that a
cluster MF measured over a limited volume would understimate the
contribution from the lowest mass stars and brown dwarfs. However,
alternative simulations of brown dwarf formation (e.g. Bate, Bonnell \&
Bromm 2003) predict that the initial velocity dispersions of low mass
stars and brown dwarfs will be similar and little mass segregation or
preferential evaporation of very low-mass objects would be expected
prior to full dynamical relaxation.

In this paper we present a new $R_{\rm c}I_{\rm c}Z$ photometric survey
of NGC 2547. This survey is deeper than that of N02,
extending to just below the substellar boundary, encompassing the ``turnover''
of the observed MF seen in the field and in the Pleiades
at $\sim 0.2\,M_{\odot}$ (see Chabrier 2003 and references
therein). The new survey also covers a wider area (0.855 square
degrees) allowing us to explore the question of mass segregation among
lower mass cluster candidates in more detail.
The observational data and analysis are presented in section~2.
The selection of candidate
cluster members and an evaluation of completeness and contamination 
are addressed in section~3. In section 4 we evaluate the
mass segregation, luminosity and mass functions
for NGC 2547. We discuss the results and draw our
conclusions in sections 5 and 6.

\section{OBSERVATIONS AND DATA ANALYSIS}

\subsection{Observations}

\label{obs}

\begin{table}
 \centering
 \begin{minipage}{140mm}
  \caption{Log of Observations.}
  \begin{tabular}{@{}cccccccc@{}}
   Field    & \multicolumn{2}{c} {Field Center (J2000)} & 
 \multicolumn{3}{c} {Exposures (s)} \\
        & R.A.       &  Dec. & $R$ & $I$ & $Z$ &  \\
        &  &  &  &  &  & & \\
01  &08 12 20.6 & -49 34 00 &  6$\times$200 & 5$\times$100 & 3$\times$300   \\
02  &08 11 16.3 & -49 34 00 &  7$\times$200 & 7$\times$100 & 3$\times$300   \\
03  &08 10 12.0 & -49 34 00 &  5$\times$200 & 5$\times$100 & 3$\times$300   \\
04  &08 09 07.7 & -49 34 00 &  6$\times$200 & 5$\times$100 & 2$\times$300   \\
06  &08 08 03.4 & -49 23 30 &  6$\times$200 & 5$\times$100 & 3$\times$300\\  
07  &08 09 07.7 & -49 23 30 &  6$\times$200 & 5$\times$100 & 3$\times$300\\ 
08  &08 10 12.0 & -49 23 30 &  6$\times$200 & 7$\times$100 & 4$\times$300\\ 
09  &08 11 16.3 & -49 23 30 &  6$\times$200 & 6$\times$100 & 3$\times$300\\
10  &08 12 20.6 & -49 23 30 &  6$\times$200 & 5$\times$100 & 3$\times$300\\
11  &08 12 20.6 & -49 13 00 &  $900+600$    & 3$\times$100 & 3$\times$300\\
12  &08 11 16.3 & -49 13 00 &  6$\times$300 & 3$\times$100 & 3$\times$300\\ 
13  &08 10 12.0 & -49 13 00 &  7$\times$200$+100$ & 3$\times$100 & 4$\times$200$+100$\\
14  &08 09 07.7 & -49 13 00 &  7$\times$200$+100$ & 3$\times$100 & 4$\times$200$+100$\\
15  &08 08 03.4 & -49 13 00 &  6$\times$200 & 5$\times$100 & 3$\times$300\\ 
16  &08 08 05.3 & -49 03 08 &  6$\times$200 & 5$\times$100 & 3$\times$300\\ 
17  &08 09 07.7 & -49 02 30 &  6$\times$200 & 5$\times$100 & 3$\times$300\\
18  &08 10 12.0 & -49 02 30 &  6$\times$200 & 5$\times$100 & 3$\times$300\\
19  &08 11 16.3 & -49 02 30 &  6$\times$200 & 5$\times$100 & 3$\times$300\\
20  &08 12 20.6 & -49 02 30 &  6$\times$200 & 5$\times$100 & 3$\times$300\\
22  &08 11 16.3 & -48 52 00 &  6$\times$200 & 5$\times$100 & 3$\times$300\\ 
23  &08 10 12.0 & -48 52 00 &  6$\times$200 & 5$\times$100 & 3$\times$300\\
24  &08 09 07.7 & -48 52 00 &  6$\times$200 & 5$\times$100 & 3$\times$300\\
51  &08 04 50.6 & -48 31 00 &  6$\times$200 & 5$\times$100 & 3$\times$300\\
52  &08 15 33.4 & -49 55 30 &  6$\times$200 & 5$\times$100 & 3$\times$300\\
\end{tabular}
\end{minipage}
\label{photlog}
\end{table}

\begin{figure}
\vspace{85mm}
\includegraphics{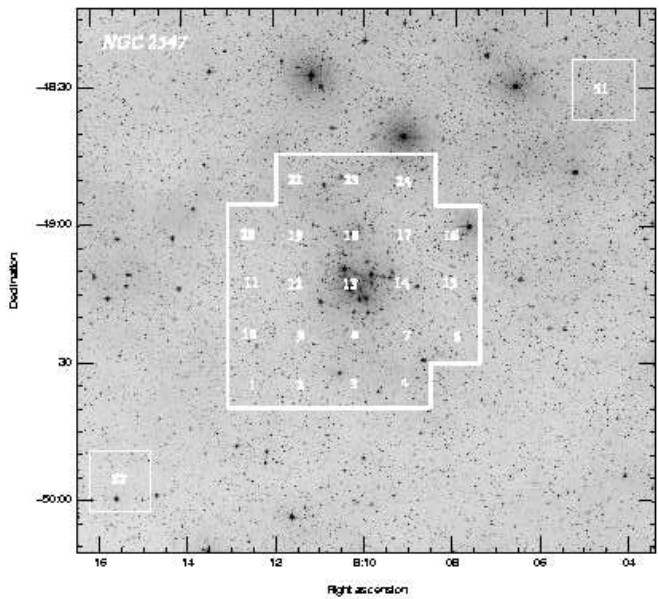}
\caption{A Digitized Sky Survey image of the region around NGC
  2547. The area surveyed in this paper are marked together with an
  approximate identification of the fields defined in Table~1.
}
\label{clusterplot}
\end{figure}

The observations were obtained between 4 February 1999 and
9 February 1999 inclusive, using a $2048\times 2048$ CCD mounted at the
Cassegrain focus of the 0.91-m telescope of the Cerro Tololo
Interamerican Observatory.  The camera had a field of
view of approximately $13.5 \times 13.5 $ square arcminutes.  
The dataset consists of multiple
exposures of 24 overlapping fields taken using each of a Harris $R$
filter, a Kron-Cousins $I$ filter and a Gunn $Z$ filter. The
approximate field centres and a log of the separate exposures obtained
are listed in Table~1.  Seeing varied between 0.9 and 2.0
arcsec in the individual frames (with a median of 1.3 arcsec and
standard deviation of 0.2 arcsec). Exposure times were adjusted a
little to achieve a reasonably uniform sensitivity across the
mosaic.  The area covered is illustrated in
Fig.~\ref{clusterplot} and covers 0.855 square degrees in total. 
Two of the individual fields are situated about a degree
from the cluster centre but along a line of constant galactic
latitude. We intended to use these fields as a check on the field
contamination present in the photometrically selected cluster candidates.

Individual images were bias subtracted (overscan subtracted and then a
median of overscan-subtracted bias frames was subtracted) and
flat-fielded using twilight sky flats from the same observing night.
The $Z$ frames were affected by fringing of night sky spectral
lines. We dealt with these by constructing a median image from all the
sky-subtracted $Z$ frames taken during the run. This ``fringe-frame''
was then multiplied by a constant and subtracted from each of the
individual $Z$ frames. The constant was modified iteratively until the
mean absolute deviation in the sky was minimised and approached the
nominal value expected from the sky signal and gain of the CCD. After
this procedure no visible trace of the fringing pattern could be
discerned.

The CCD was affected by a number of bad columns and pixels. We also
found that occasional images were affected by 27 bad rows at the bottom
of the CCD. Bad pixel masks were constructed for all frames and used in
the subsequent reduction. The bad pixels were replaced with the median
of surrounding pixels. It is worth emphasizing that these patched
pixels were used in the initial source searching of the summed frames,
but the final combined photometry uses only stellar images with no
patched pixels where possible.

\subsection{Photometric calibration}

\label{standards}

Photometric calibration was achieved with reference to
measurements of Landolt (1992) standard fields taken on a
number of occasions on each night. The flux from standard stars was
measured inside a 6 arcsecond radius aperture and sky subtracted using
the median sky measured in a much larger annulus.

The $I_{\rm c}$ and $R_{\rm c}-I_{\rm c}$ values listed in
Landolt (1992) were used to find a weighted least squares solution to the
following equations on each night
\begin{equation}
I_{\rm c} = i + c_{ri} (R_{\rm c} - I_{\rm c}) - k_{i}X + z_{i}\, ,
\label{ieq}
\end{equation}
\begin{equation}
R_{\rm c} - I_{\rm c} = a(r - i) - k_{ri}X + z_{ri}\, ,
\label{rieq}
\end{equation}
where $i$ and $r$ are instrumental magnitudes (corrected for exposure
time), $k_{i}$ and $k_{ri}$ are extinction coefficients, $X$ is the
airmass of an observation, $z_{i}$ and
$z_{ri}$ are photometric zeropoints, $c_{ri}$ is a
colour term to take into account any mismatch between the $I$
filter we used and the Cousins system, and $a$ is a similar term for
the $R_{\rm c}-I_{\rm c}$ colour.

For our best photometric night (7 February 1999) we obtained a solution
with $c_{ri}=0.074$, $k_{i}=0.237$, $a=0.975$, $k_{ri}=0.090$. The rms
discrepancy from this fit using 26 points was 0.04 mag in $I_{\rm c}$ and
0.02 mag in $R_{\rm c}-I_{\rm c}$. The addition of about 0.02 mag in
quadrature to the instrumental $i$ and $r-i$ uncertainties leads to a
reduced chi-squared of 1. However, only two standards with $R_{\rm c} -
I_{\rm c}>1.5$ were measured and one of these (SA\,98-L5) has quite
uncertain colours.  Systematic errors of up
to 0.1 mag in the colours and magnitudes cannot be ruled out for stars with $R_{\rm
c}-I_{\rm c}>1.5$, although it is encouraging that the colour terms in
equations~\ref{ieq} and \ref{rieq} are small and the residuals to the
fits at $R_{\rm c} - I_{\rm c}<1.5$ do not suggest any non-linearities
in the calibration.

There is no equivalent of the Landolt standards for the $I_{\rm c}-Z$
colour. Instead we used observations of stars in Landolt field SA\,98
which have had their $I_{\rm c}-Z$ colours defined by Zapatero-Osorio et
al. (1999), along with two early A stars from Landolt field SA\,99 for
which we assumed that $I_{\rm c}-Z=R_{\rm c}-I_{\rm c}$. Using an equation of the form
\begin{equation}
I_{\rm c} - Z = b(i - z) - k_{iz}X + z_{iz}\, ,
\label{izeq}
\end{equation}
we found $b=0.85$ and $k_{iz}=0.07$, with an rms discrepancy of less
than 0.01 mag from 9 points.

These coefficients were applied to 5 of our fields which were taken on
the same night and in an airmass range comparable to the standards
(fields 1, 2, 6, 7, 10) . These fields were considered ``photometric''
in a normalisation procedure that uses the considerable (2-3 arcminute)
overlaps between the fields (described below and in more detail by
N02). The remaining fields were put onto the standard
system using the same coefficients, but their zeropoints were allowed
to vary in the normalisation procedure.

\subsection{Photometry and astrometry in the cluster}

\label{photastrom}

From the processed CCD frames the analysis proceeded using the {\sc
cluster} software suite described in detail by N02.  In brief, this
consisted of searching for objects in the summed I-band image of each
field. A signal-to-noise ratio threshold of 5 was used for the search.
Shifts between each frame and the summed I-band image were
parameterised with a 6-coefficient solution, and optimal photometry
(see Naylor 1998) of each object performed in each individual frame at
fixed positions determined from the summed I-band image.  Bright,
unsaturated stars in each frame were used to derive the ``profile
correction'', which was applied to the optimal photometry values to
place them onto the system defined by the aperture photometry of
standard stars. The profile correction was successfully modelled as a
quadratic function of the horizontal and vertical coordinates in each
frame.

Profile-corrected photometry measurements were adjusted for differences
in the airmass of each frame, weighted by their statistical errors and
combined. Additional, magnitude-independent errors of between 0.004 and
0.010 mag were added (in quadrature) to the results for each individual
frame to yield a plot of reduced chi-squared versus signal-to-noise
ratio for each field which had a uniform value of 1 (see
N02 for details).  Instrumental magnitudes were combined with the
average airmass of each set of observations and equations~\ref{ieq},
\ref{rieq} and~\ref{izeq} used to obtain preliminary magnitudes and
colours for stars in each field.

Astrometric calibration was defined by approximately 1000 objects in
each frame that were identified on UK Schmidt $J$ plates as part of the
SuperCOSMOS sky survey (Hambly et al. 2001a).  The rms discrepancy in
position for each identified object, after correction for proper
motions in the SuperCOSMOS catalogue, was 0.20--0.25 arcseconds. This
uncertainty is mainly in the positions from the SuperCOSMOS catalogue
-- comparison of positions for stars in our final catalogue with those
common to the NGC 2547 catalogue presented by N02, which was calibrated
using the same SuperCOSMOS data, indicate an internal astrometric
precision for each catalogue of 0.05 arcseconds (rms) in each
coordinate.

The overlapping regions between fields were used to refine the
photometry and produce a uniformly calibrated mosaic. A shift was
applied to the colours and magnitudes of each field which minimised the
average discrepancies of all the stars in the overlap regions, subject
to the constraint that the average shift applied to all those fields
considered ``photometric'' (see section~\ref{standards}) was
zero. After this procedure the rms discrepancy in the photometry for
each field was 0.010 mag in $I$, 0.007 mag in $R_{\rm c}-I_{\rm c}$ and
0.009 mag in $I_{\rm c}-Z$. The final catalogue was produced by
averaging the results for stars found in more than one field.

There were several departures from the procedures described by N02
which are worth noting. 
\begin{enumerate}
\item In N02 the source searching algorithm looked for ``islands'' of
  pixels which were some multiple of the noise above the sky
  level. This threshold was iteratively reduced by factors of 4 in
  order to isolate stars situated in the wings of other stars, where
  the minimum flux at a point between them may not fall below the
  original threshold. It became apparent during the reduction of this
  dataset that a better thresholding step is a factor of $\surd{2}$. In
  the case of sky-limited, Gaussian stars, this ensures that stars
  which are formally resolved (i.e. their flux profiles overlap at a
  point beyond their FWHM) are clearly separated as two sources by our
  search technique in the course of at least one iteration. Though this
  choice slows down the source searching procedure it is far from being
  the rate-determining step in the overall data reduction process.
\item The photometry in the catalogue now uses the character-based 
  flagging scheme described in Burningham et al. (2003) to identify
  non-stellar objects, stars lying on bad pixels, poor sky estimation etc.
\item In our reduction scheme, ``non-stellar'' means that the optimal
photometry technique will not yield a reliable measurement because the
object is significantly extended, confused with another object or
affected by the flux from a nearby brighter star. We found that testing
for non-stellarity only in the summed I-band image was
insufficient. Faint, very red, low-mass cluster candidates might appear
isolated in the $I$ band, but their $R$ band flux could be biased
upward by a nearby, unrelated field star with a bluer spectrum. Hence
the estimated $R_{\rm c}-I_{\rm c}$ colour of the cluster candidate
could be significantly underestimated.  To counter this, an additional
test for non-stellarity was made in the summed $R$ band image. If
either test resulted in a non-stellar classification then this was
reflected in the flags of all three magnitudes and colours. It was not
necessary to extend the test to a summed Z-band image because the vast
majority of stars (see Figs.~\ref{cmdri} and \ref{cmdiz}) are bluer
than the faint cluster candidates.

\item The sky estimation technique (see Naylor 1998) is robust to
  nearby stars inside the ``sky box'' or smooth gradients
  in the sky background. However, curvature in the background, such as
  for objects situated in the wings of very bright or saturated stars
  will result in a histogram of pixel values that is highly asymmetric
  or not well fit by the skewed Gaussian model we adopt. To avoid
  spurious detections or significantly mis-estimated sky backgrounds,
  we demand that the reduced chi-squared of the fit to the sky
  histogram is less than 3 {\it and} that the skewness parameter is
  less than 0.3. If either test is failed then an ``ill-determined sky'' flag is set.

\end{enumerate}

\subsection{The final catalogue}
\label{finalcat}

The final catalogue contains 133788 stars, of which 49223 have one or
more quality flags set. It is available in electronic format 
and is described in Appendix~A.

Before using the catalogue it is worth examining sources of
internal and external errors in the catalogue measurements. 
\begin{enumerate}
\item
The astrometry has an internal precision of about 0.05 arcsec (1 sigma)
in each coordinate for objects with a signal-to-noise ratio (in the $I$ band) of $>10$.
The systematic external uncertainties are equivalent to the external uncertainties
present in the SuperCOSMOS astrometric calibration (about $\pm0.2$
arcsec compared with the International Celestial Reference Frame --
Hambly et al. 2001b).
\item The uncertainties for the magnitudes and colours reflect
  a combination of the statistical uncertainties of the optimal
  photometry flux measurements 
   with an additional, small, magnitude-independent error added to
  each frame (see section~\ref{photastrom}).
\item The catalogue {\it does not} include uncertainties revealed
  by the rms discrepancies in the zeropoints of the overlapping regions
  between fields (see section~\ref{photastrom}). These 
uncertainties should be added to the errors presented in the final
catalogue if stars are to be compared or combined across different
fields or the survey as a whole (as is the case in this paper).
\item Finally there are systematic external photometric uncertainties,
  which are generally 
  less than 0.02 mag but perhaps as large as 0.1 mag for the reddest stars (see
  section~\ref{standards} and below).
These may be important when comparing with other catalogues or with
theoretical models.

As a check on external photometric precision we compared our $I_{\rm c}$
photometry with that in N02. A subset of stars 
with no quality flags set
and $I_{\rm c}$ errors less than 0.1 mag was selected from
each catalogue. These subsets were cross-correlated 
using a 1.0 arcsec correlation radius, finding 13032 matches.
We find the following weighted least-squares fit relationships after
adding the small errors revealed by the rms discrepancies in the
zeropoints of the overlapping regions {\em in each} dataset: 
\bd
I_{\rm c} - I_{\rm N02} = -0.012(3) + 0.0019(2)\,I_{\rm c}\, , 
\ed
\bd
I_{\rm c} - I_{\rm N02} = -0.004(9) + 0.043(2)\,(R_{\rm c}-I_{\rm c})\, ,
\ed
for $12<I_{\rm c}<20$ and $0.1<R_{\rm c}-I_{\rm c}<2.0$, and where the
numbers in brackets are 1-sigma uncertainties in the final significant
figure. Higher order polynomials provide no significantly better fits.
One of the strengths we claim for our techniques is the complete
trustworthiness of the internal photometry errors (for unflagged
objects). The reduced chi-squared values for the above fits, of 1.37
and 1.22 respectively, provide reassurance that this is the case,
especially given the likelihood that at least some stars are truly
variable.  There is a significant difference and trend with colour
between the N02 and our values -- the N02 $I_{\rm c}$ values are about
0.08 mag brighter at $R_{\rm c}-I_{\rm c}=2$. As both datasets suffer
from the same lack of very red calibration stars it would be unwise to
attach more weight to one than the other.

\end{enumerate}

\section{Cluster Membership}

\subsection{Colour magnitude diagrams}
\label{cmds}

\begin{figure}
\vspace{70mm}
\includegraphics{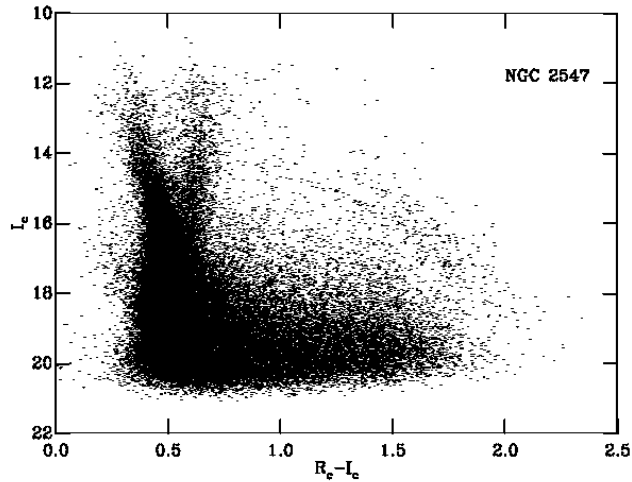}
\caption{The $I_{\rm}$ vs $R_{\rm c}-I_{\rm c}$ CMD for the entire
  catalogued area of NGC 2547. Only unflagged objects with
  signal-to-noise ratio better than 10 are shown.
}
\label{cmdri}
\end{figure}
\begin{figure}
\vspace{70mm}
\includegraphics{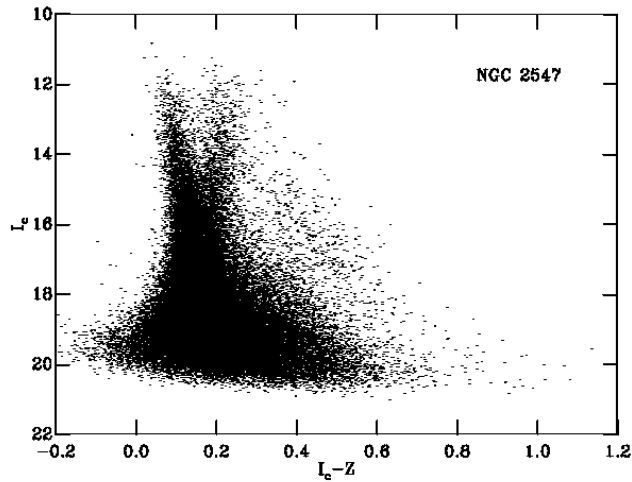}
\caption{The $I_{\rm}$ vs $I_{\rm c}-Z$ CMD for the entire
  catalogued area of NGC 2547. Only unflagged objects with
  signal-to-noise ratio better than 10 are shown.
}
\label{cmdiz}
\end{figure}

Figures~\ref{cmdri} and \ref{cmdiz} show the $I_{\rm c}$ vs $R_{\rm
  c}-I_{\rm c}$ and $I_{\rm c}$ vs $I_{\rm c}-Z$ colour-magnitude
diagrams (CMDs) for all unflagged stars with a signal-to-noise ratio (in
colour and magnitude) better than 10 (62722 and 55174 stars respectively).

The NGC 2547 PMS is clearly visible in both diagrams. It is better
delineated when looking at the positions of X-ray active stars (see
section~\ref{isochrone} and N02), but we show these diagrams to make
obvious the concentration of cluster stars (especially in
Fig.~\ref{cmdri} for $0.9<R_{\rm c}-I_{\rm c}<2.0$) and their
separation from the contaminating background.

\subsection{Model Isochrones}

\label{isochrone}

\begin{figure}
\vspace{70mm}
\includegraphics{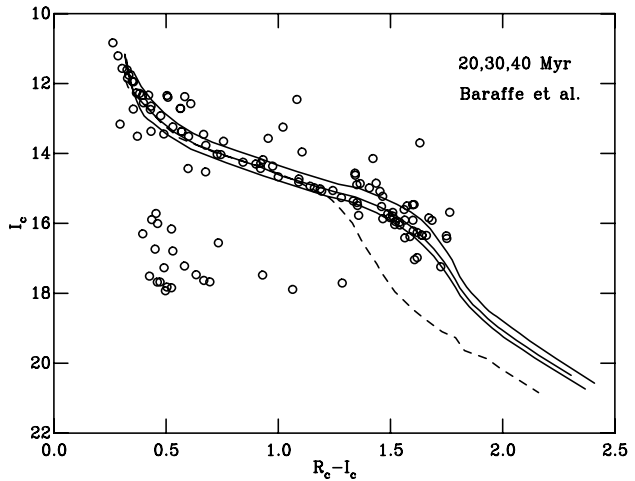}
\caption{X-ray selected objects in the field of NGC 2547 with 20, 30 and
  40\,Myr empirical isochrones generated from the B02
  models at an assumed distance modulus and reddening of 8.1 and
  $E(R-I)=0.043$. The dashed line indicates a 30\,Myr isochrone taken
  directly from the B02 models. 
}
\label{xrayricmd1}
\end{figure}

\begin{figure}
\vspace{70mm}
\includegraphics{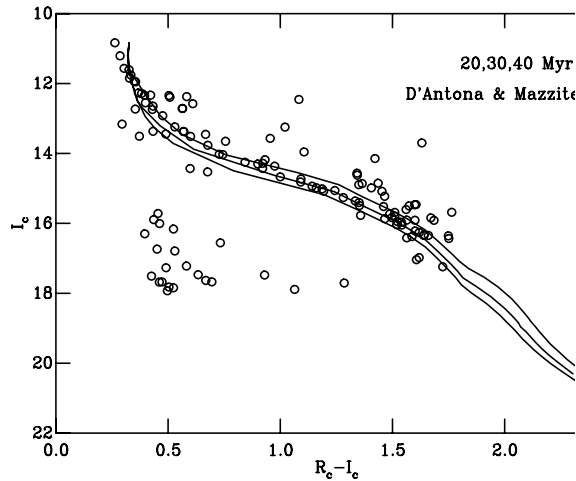}
\caption{As for Fig.~\ref{xrayricmd1}, but using the DAM97 models.
}
\label{xrayricmd2}
\end{figure}

\begin{figure}
\vspace{70mm}
\includegraphics{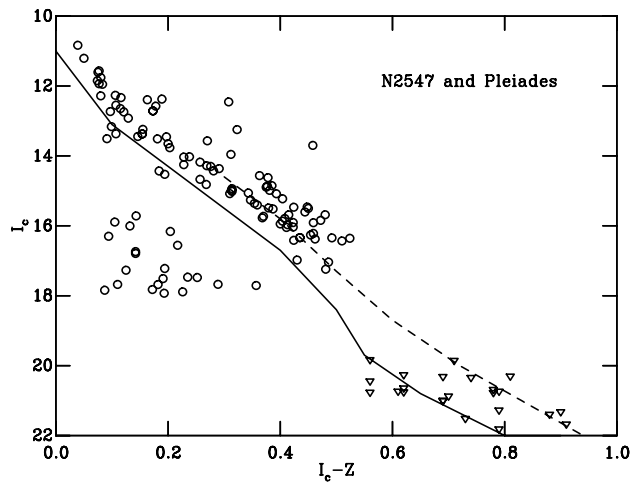}
\caption{The $I_{\rm}$ vs $I_{\rm c}-Z$ CMD for the X-ray selected
  targets in NGC 2547 (circles) plus Pleiades BDs from Zapatero et
  al. (1999) shifted to the same distance as NGC 2547 (triangles). The
  solid line indicates the lower envelope used in our membership
  selection criteria for NGC 2547. The dashed line is the PMS locus we
  have assumed for NGC 2547 when generating a simulated catalogue (see section~\ref{complete}).
}
\label{xrayizcmd}
\end{figure}

To select cluster members and assign masses to
stars on the basis of their colours and magnitudes, we must choose an
isochrone that adequately models the data. To aid 
this process we fit a subset of the catalogue 
identified as optical counterparts to {\em XMM-Newton} X-ray sources.
The rationale is that young stars in NGC 2547 are known to be very
X-ray active, typically orders of magnitude more active than
contaminating field stars at a similar position in the observed CMDs
(Jeffries \& Tolley 1998). An X-ray selected sample should present a
relatively clean PMS sample with which to fit model isochrones.  The details
of the {\em XMM-Newton} observations will be presented elsewhere
(Jeffries et al. in preparation). In brief, they consist of a 50\,ks
pointing at the centre of the cluster, from which a total of 213
significant X-ray sources were found in one or a combination of the two
{\em PN} plus {\em EPIC} detectors and within a 17 arcminute radius of
the {\it XMM-Newton} field centre.
These sources were correlated with our photometric catalogue, but to
to minimize the number of spurious correlations, we restricted the catalogue to
$I_{\rm c}<18$, because experiments showed that no
plausible PMS cluster members were X-ray detected at lower
luminosities. The X-ray optical counterparts are displayed in
Figures~\ref{xrayricmd1}, \ref{xrayricmd2} and \ref{xrayizcmd}.
The objects in the lower left of these plots are not cluster members
and are most likely extragalactic X-ray sources that are randomly
correlated with faint objects in the bulk of the CMD background population.


To select appropriate isochrones we used the models of D'Antona \&
Mazzitelli (1997, DM97) and Baraffe et al. (2002, B02 -- with mixing
length parameter set to 1.9 scale heights for masses greater than
0.6$M_{\odot}$). The B02 models predict optical and infrared
magnitudes as a function of age and mass, using the same model
atmospheres that provide boundary conditions for the evolutionary
model. However, it is well known that there are problems (mainly at $V$
and $R$) with these magnitudes, probably due to omissions in the
various sources of opacity at these wavelengths (see Baraffe et
al. 1998). In Figure~\ref{xrayricmd1} we show a model 30\,Myr B02
isochrone (as a dashed line) constructed assuming an intrinsic distance modulus of
8.1 - clearly it is not a reasonable match to the data, being far too
blue for the fainter X-ray sources.

Instead we adopt the approach of Jeffries, Thurston \& Hambly (2001)
and N02. We assume that the large amount of excellent photometric data
for the Pleiades {\em defines} an empirical isochrone at an age of
$\simeq 120$\,Myr\footnote{We use $RI$ data for confirmed Pleiades
  proper-motion members from Stauffer (1982), 
  Stauffer et al. (1984), Bouvier et
  al. (1998) and Moraux, Bouvier \& Stauffer (2001).  Where necessary,
  Kron-Cousins photometry is transformed to Cousins photometry using
  the equations in Bessell \& Weis (1987)}.  Combined with empirical
$I_{\rm c}$-band bolometric-correction versus $R_{\rm c}-I_{\rm c}$
relationships (a quadratic fit to data from Leggett 1992 and Leggett et
al. 1996), this empirical isochrone is used to determine a
colour-$T_{\rm eff}$ relation that can be applied to isochrones at any
age.  Isochrones are generated from each set of evolutionary models,
converted to colours and magnitudes and adjusted for the distance
modulus, extinction and reddening for NGC 2547. The data here cannot
determine any of these parameters independently of age, so we assume an
intrinsic distance modulus of 8.1 (see N02), visual extinction, $A_{\rm
  V}$, of 0.19 magnitudes (Clari\'a 1982) and $E(R_{\rm c}-I_{\rm c})$
of 0.043.

$I_{\rm c}$ versus $R_{\rm c}-I_{\rm c}$ isochrones are shown in
Figs.~\ref{xrayricmd1} and~\ref{xrayricmd2} for the B02 and DM97 models
respectively. The isochrones are a good fit to the X-ray selected
cluster members at ages of 35\,Myr (B02) and 30\,Myr (DM97) for the
range of colours of low mass objects discussed in this paper
($R_{\rm c}-I_{\rm c}>0.7$). The age uncertainties are dominated by
uncertainties in the cluster distance modulus -- a plausible $\pm 0.1$
error in distance modulus leads to a $\mp5$\,Myr age change.  There are
significant differences in the shapes of the B02 and DM97 isochrones
for $I_{\rm c}>17$. In principle these differences could be tested by
firm membership information for fainter cluster candidates. In practice
we need to perform our analysis for both isochrones in terms of
selection of candidate members {\em and} subsequent calculation of MFs etc.

The same approach cannot be adopted for the $I_{\rm c}$ versus $I_{\rm
c}-Z$ CMD where there is insufficient, consistently calibrated Pleiades
photometry from which to define the $I_{\rm c}-Z$ versus $T_{\rm eff}$
relation. Instead, for $I_{\rm c}<17.5$ we define a locus that forms a
conservative lower boundary to the X-ray selected stars and for fainter
stars we extend this boundary by assuming that NGC 2547 candidates must
lie above a set of consistently calibrated Pleiades low-mass stars
(from Zapatero-Osorio et al. 1999). Our X-ray selected stars, the Pleiades
objects (shifted to the same distance modulus as NGC 2547) and our
defined lower boundary are shown in Fig.~\ref{xrayizcmd}. The point of
inflexion at $I_{\rm c}-Z\simeq0.5$ {\em does} appear to be a genuine
feature of the models and filter bandpasses (e.g. Moraux et
al. 2003).

\subsection{Selection criteria}
\label{select}

\begin{figure}
\vspace{70mm}
\includegraphics{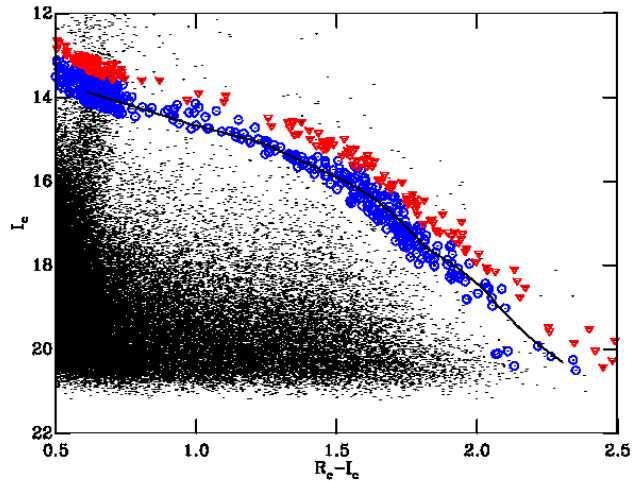}
\caption{The selected stars from our total catalogue. Circles are
  ``single'' stars, triangles are possible unresolved, high mass-ratio
  binary systems. The solid line is the 30\,Myr DM97
  isochrone used in the selection procedure.
}
\label{selectri}
\end{figure}
\begin{figure}
\vspace{70mm}
\includegraphics{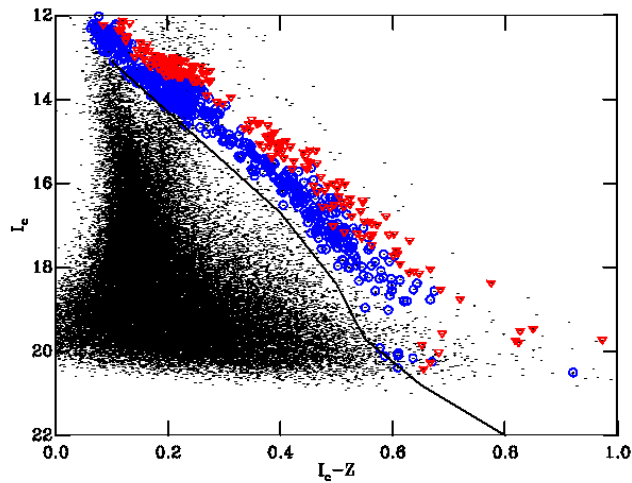}
\caption{The selected stars from our total catalogue. Symbols as for
  Fig.~\ref{selectri}. The solid line is the locus above which our
  candidates are required to lie.
}
\label{selectiz}
\end{figure}
\begin{figure}
\vspace{70mm}
\includegraphics{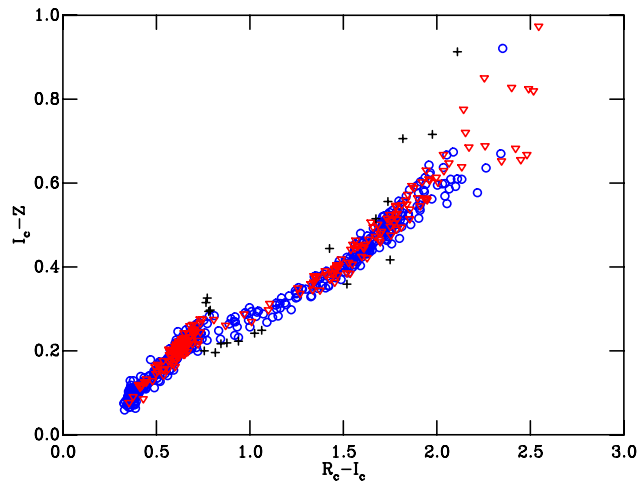}
\caption{The selected stars from our total catalogue. Symbols as for
  Fig.~\ref{selectri} and crosses indicate those stars which passed
  selection on both CMDs but which have positions in this plot that are
  $>4$ times their error bar away from the mean trend defined by
  all other cluster candidates. The abrupt change in slope at
  $R_{\rm c}-I_{\rm c}\leq 0.7$ is probably due to heavy
  contamination by background giants.
}
\label{selectriiz}
\end{figure}

Selection of candidate members in the CMDs is an arbitrary
process to some extent. As our aims are to produce luminosity and
mass functions (LFs and MFs) we choose to err on the side of generosity, defining
tests that all members {\em should} pass (although see
section~\ref{complete}), at the expense of including more contaminating
field objects (see section~\ref{contaminate}).
For each of the two sets of model isochrones in the $I_{\rm c}$ vs
$R_{\rm c}-I_{\rm c}$ CMD we select cluster candidates according to the
following tests.
\begin{enumerate}
\item The star must be unflagged and have photometric errors less than
  0.1 mag in $I_{\rm c}$ and $I_{\rm c}-Z$ and less than 0.2 mag in
  $R_{\rm c}-I_{\rm c}$. The less stringent constraint on $R_{\rm c}-I_{\rm c}$
  is justified because it is approximately twice as sensitive to
  effective temperature as $I_{\rm c}-Z$.
\item The star must be redder in $I_{\rm c}-Z$ than the boundary
  defined by the solid line in Fig.~\ref{xrayizcmd}.
\item The star must lie between 0.9 mag above and 0.25 mag below the
  chosen isochrone in Figs.~\ref{xrayricmd1} and ~\ref{xrayricmd2}. In
  addition we allow a further degree of freedom by adding the
  photometric error (the sum of uncertainties in the colour
  and magnitude after allowing for the gradient of the
  isochrone). Furthermore for stars with $R_{\rm c}-I_{\rm c}>1.5$ we
  allow another $\pm0.05$ mag of possible error (in $I_{\rm c}$ and
  $R_{\rm c}-I_{\rm c}$) to allow for plausible uncertainties
  in the photometric calibration.
\item For stars that pass the above tests there is still the
  possibility that a star's $I_{\rm c}-Z$ and $R_{\rm c}-I_{\rm c}$
  may be incompatible. Although we do not have a good estimate
  of the intrinsic $I_{\rm c}-Z$ locus at faint magnitudes, we iteratively
  reject stars that lie more than 4$\sigma$ from a cubic
  fit to candidate members in the $I_{\rm c}-Z$ vs $R_{\rm c}-I_{\rm
  c}$ plot, where $\sigma$ is the appropriately combined photometric
  uncertainty in both colours. This test is only applied to stars with
  $R_{\rm c}-I_{\rm c}>0.7$ because the sample is dominated at bluer
  colours by background giants (see section~\ref{contaminate}).
\item Finally, those stars that pass all tests {\em and} lie more
  than 0.5 mag above the $I_{\rm c}$ vs $R_{\rm c}-I_{\rm c}$ isochrone
  are classified as {\em possible} binary systems with large mass ratios.
  The threshold of 0.5 mag corresponds roughly to mass ratios,
  $q\geq0.35-0.65$ (see section~\ref{binary}).
\end{enumerate}

Figures~\ref{selectri}, \ref{selectiz} and \ref{selectriiz} show the
results of this selection procedure in the case of the DM97 isochrone
in Fig.~\ref{xrayricmd2}.  The selected candidate members are given in
Tables~A2 and~A3 which are available in electronic format -- see Appendix~A.

\subsection{Completeness}
\label{complete}

\begin{figure}
\vspace{70mm}
\includegraphics{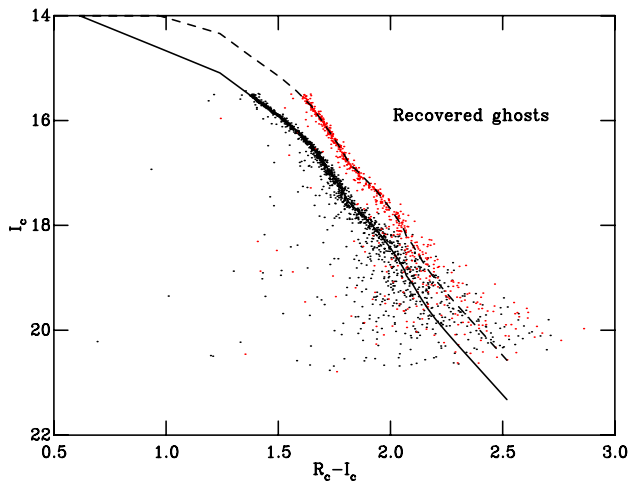}
\caption{The $RI$ CMD of ghost stars recovered by our detection and
  photometry procedures {\em and} which are unflagged and pass the
  necessary signal-to-noise thresholds for inclusion as canididate
  members. The solid and dashed lines show the
  injected single star and binary sequences.
}
\label{ghostricmd}
\end{figure}
\begin{figure}
\vspace{70mm}
\includegraphics{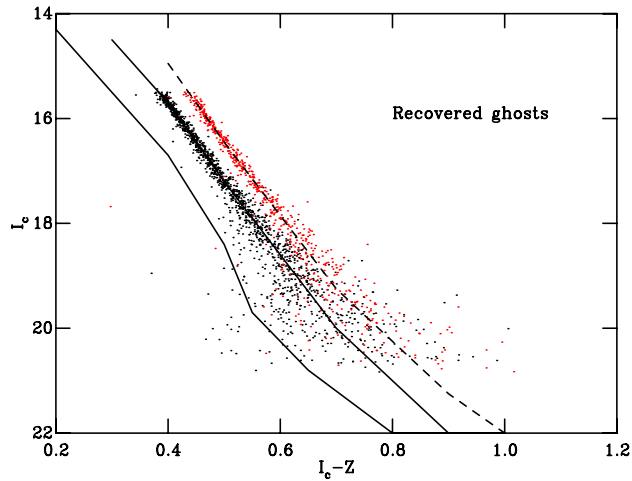}
\caption{As for Fig.~\ref{ghostricmd} but for the $IZ$ CMD.
The lower of the two solid line indicates the lower envelope for
candidate membership in this diagram (see Fig.~\ref{selectiz})
}
\label{ghostizcmd}
\end{figure}
\begin{figure}
\vspace{140mm}
\includegraphics{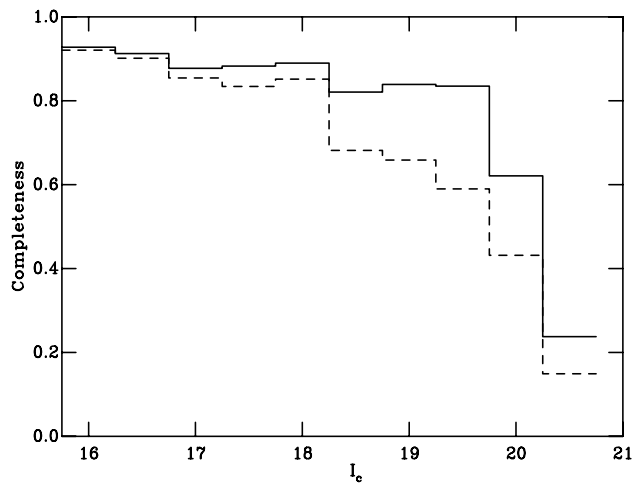}
\includegraphics{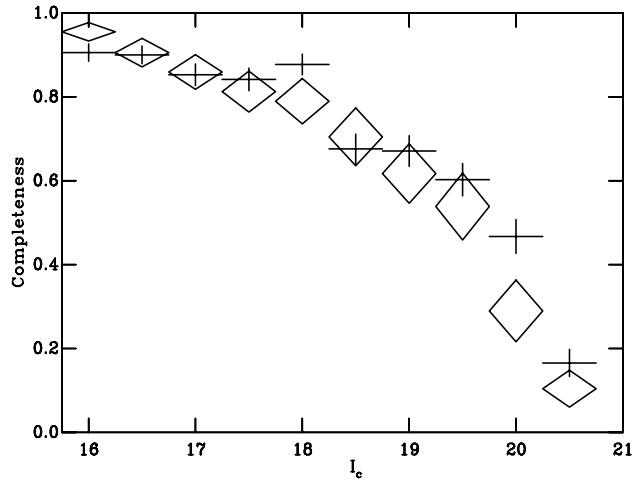}
\caption{(Top) The CF determined either by the fraction of stars that
  are detected, unflagged and of sufficient signal-to-noise ratio to
  enter our candidate member catalogue (solid line) or the fraction of
  stars that pass {\em all} the membership criteria (dashed
  line). (Bottom) A comparison of this latter definition of CF for the
  inner 15 arcminute radius subset of our survey (error diamonds) with the
  CF for the area outside this (error crosses). Note that uncertainties have been
  estimated using the binomial distribution. 
}
\label{completefig}
\end{figure}

Completeness is an important issue when deriving LFs and MFs in
clusters and is rarely addressed comprehensively in the literature.
There are several reasons why members of the NGC 2547 cluster may not
appear in our final lists of candidate members. (a) They may not be
detected at all, either because they are too faint or too close to
another star. (b) They may be flagged as non-stellar or fall on bad
pixels or be otherwise flagged as having poor quality photometry. (c) Their
photometry may scatter outside the selection criteria defined in the
previous section. It is not sufficient to simply decide upon a
magnitude threshold above which one expects to merely {\em detect} some
fraction of stars.

To estimate completeness we generated
a catalogue of ``ghost'' stars (see N02), that were scaled
copies (with the appropriate noise characteristics) of bright,
unflagged stars in the total catalogue. The colours and magnitudes of
the ghosts were allocated according to the isochrones defined for
$I_{\rm c}$ vs $R_{\rm c}-I_{\rm c}$ in section~\ref{isochrone}. For
$I_{\rm c}-Z$ we used the isochrone defined in
Figure~\ref{xrayizcmd}. It is important to provide appropriate
colours for the ghosts, as completeness will
certainly be colour-dependent.  It is impossible to
precisely simulate the properties (LF, binary frequency, binary mass
ratio distribution) of the cluster without knowing what they are {\em a
priori}. We made the simplification that the cluster can be represented
with a single star population and a binary star population with unit
mass ratio.  Because we allowed quite generous bounds on our
selection criteria in section~\ref{select}, this simplification should
not have any great influence on the derived completeness information.

Two simulations were run. The first contained 2013 ghost stars 
placed on the single star cluster isochrones between $15.75<I_{\rm
  c}<22.0$ (we used the DM97
$I_{\rm c}$ vs $R_{\rm c}-I_{\rm c}$ isochrone), with a declining
LF (LF\,$\propto \exp[-I_{\rm c}/4]$) that roughly mimics that seen in
the data (see section~\ref{lf}), although with a
shallower gradient\footnote{We anticipated that incompleteness would
grow with magnitude, so this choice was approximately consistent with the
measured LF. N02 explain why an appropriate choice of
LF is important.}. The second contained 862 ghost stars placed on a
binary sequence, 0.75 magnitudes above the single star locus, covering
the same magnitude range and with the same LF. This binary frequency
(number of binary systems divided by total number of systems) of 30 per
cent is about the average frequency of binary systems found (using the
definition in section~\ref{select}) in our data with
$R_{\rm c}-I_{\rm c}>0.8$.

The ghosts were placed in {\em individual} image frames according to
their positions in the catalogue. The spatial distribution of the
ghosts was tied to the parents from which they were scaled and was
approximately uniform. The individual frames were summed and
searched and pass through {\em precisely} the same operations
as the genuine data. The number of ghosts was only a
small perturbation (typically $\sim100$ ghosts in a frame containing
8000 other stars) on the detection efficiency. They did not
significantly increase the crowding in the frame and rarely did
ghosts interfere with the detection and photometry of another ghost.
However, the number of ghosts was large enough compared with the numbers
of candidate NGC 2547 members, that statistical
uncertainties in completeness are unimportant in our final derived
 LFs and MFs.

CMDs for recovered ghosts that were unflagged and have the required
signal-to-noise-ratios are shown in Figs.~\ref{ghostricmd}
and~\ref{ghostizcmd}. Some ghosts (a few per cent) were found with
colours and magnitudes completely different to those simulated. We
examined a number of these pathological objects and found that these
were instances where the ghost was placed almost directly on top of an
already existing star and was not flagged as non-stellar.  As most
stars in our fields are bluer than the cluster sequence, the photometry
of these objects tended to be bluer than expected.  This could happen
to {\em real} cluster stars, so it is appropriate that these are
included in completeness considerations.  It is also clear that in the
absence of detailed {\em a priori} knowledge of the mass-ratio
distribution, any attempt to determine binary frequencies based on
position in the CMDs will be compromised by scatter for $I_{\rm
c}>18.5$ ($R_{\rm c}-I_{\rm c}>2.0$).

We determined two ``completeness functions'' (CFs) using these
simulations. The first is the percentage of ghosts as a function of
$I_{\rm c}$ that were detected, unflagged and satisfied our
signal-to-noise criteria in section~\ref{select}.  The second took the
recovered ghosts through the full membership selection procedure,
discarding those with colours and magnitudes inconsistent with membership.
This second CF is appropriate for correcting the observed
cluster LFs and MFs.  Figure~\ref{completefig} shows the results.
The upper plot (for the whole surveyed area)
demonstrates that the second definition of CF results in significantly
lower completeness at fainter magnitudes -- a difference that could
change the slope of a completeness-corrected LF or MF.

Both CFs were investigated for any spatial variation. It could be the
case that the presence of more bright stars and hence higher and more
complicated background in the cluster centre could compromise
sensitivity or photometric accuracy.  We derived CFs for the inner 15
arcminute radius (using the cluster centre defined by Littlefair et
al. 2003) and compared them to CFs outside this area and these are
shown in the lower plot of Fig.~\ref{completefig}. There is some
evidence that the inner CF is lower for $I_{\rm c}>19.5$, but not for
brighter stars.  In subsequent derivations of MFs and LFs we use a
linear interpolation of the appropriate CF, depending on whether stars
are inside or outside this inner circle. We will also assume that the
CF for stars with $I_{\rm c}<15.75$ is 95 per cent.

\subsection{Contamination}
\label{contaminate}

\begin{figure}
\vspace{70mm}
\includegraphics{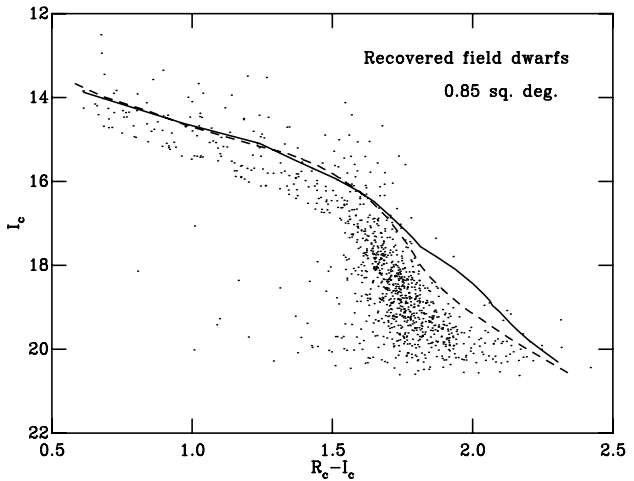}
\caption{The recovered CMD for simulated field dwarfs (see
  section~\ref{contaminate}) for a survey area of 0.855 square degrees and
  including stars out to a distance of 1200\,pc. The solid and dashed
  lines show the DM97 and B02
  isochrones that were used to select members in
  section~\ref{select}. Only unflagged stars with a signal-to-noise
  ratio sufficient for membership status are shown.
}
\label{contamfig}
\end{figure}
\begin{figure}
\vspace{70mm}
\includegraphics{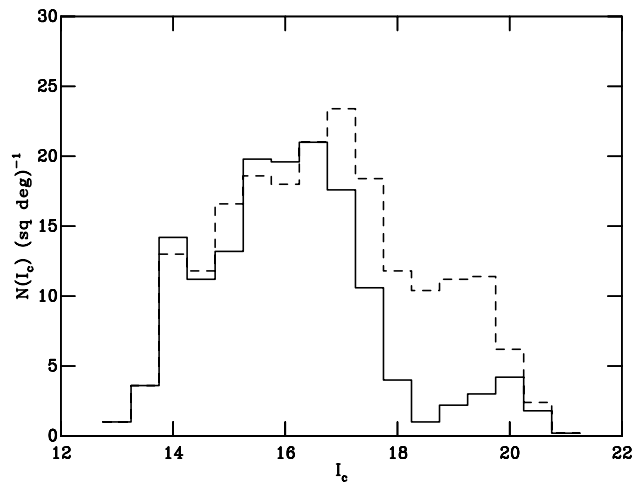}
\caption{The luminosity functions (LFs) for contaminating field
  stars that have colours and magnitudes that would satisfy our
  membership selection criteria 
  using the DM97 or B02
  isochrones (solid and dashed lines) respectively.
}
\label{contamlf}
\end{figure}

Another issue hindering accurate reconstruction of LFs and MFs
is contamination of CMDs by objects unassociated with the
cluster.  Our photometrically selected sample of cluster candidates
contains contaminants, consisting of background giants and
foreground dwarfs.  Background G/K giants can be seen in
Fig.~\ref{cmdri} as a dense ``finger'' of contamination dominating
the cluster sequence at $0.5<R_{\rm c}-I_{\rm c}<0.7$. It is doubtful
that evolved stars make much contribution to contamination outside of
these colours. The cluster is at galactic latitude -8.6 degrees
and longitude 264.5 degrees. A background M giant/supergiant with
$R_{\rm c}-I_{\rm c}\geq 1.0$ would have an absolute $I$ magnitude
less than -2 and in order to have $I_{\rm c}\geq 15$ and
encroach on the NGC 2547 PMS would be at a distance $\geq 25$\,kpc and
be more than 3.7\,kpc out of the plane. The space density of such objects is
essentially zero more than 2\,kpc above the plane (Branham 2003) and so
if we restrict our analysis to stars with $R_{\rm c}-I_{\rm c}>0.7$
then we need not consider contamination by background giants any
further.

Contamination by older foreground field dwarfs with similar colours but
lower luminosities than the NGC 2547 PMS stars is a more serious
problem. The separation of the PMS from the bulk concentration of field
objects in the CMDs is encouraging, but a quantitative estimate is
desirable. We simulate the field star population using
the $I$-band luminosity function of stars within 8\,pc (taken from Reid
\& Hawley 2000). Taking this local LF (in stars\,pc$^{-3}$\,mag$^{-1}$)
as a weighting function, we assign absolute $I$ magnitudes randomly to
a large number of stars. The local LF contains many resolved binary
systems, the vast majority of which would be unresolved in our
survey. To account for this we assume 75 per cent of stars are
part of binary systems (the fraction found in the 8\,pc sample),
consisting of two stars drawn randomly from the local LF.  We further
assume that the spatial density varies as $N_{0}\exp(-z/h)$, where $z$
is the height below the galactic plane, $h=270$\,pc is the scale
height for low-mass stars (see Kroupa, Tout \& Gilmore 1993) and 
$N_{0}$ is the density implied by the local LF. We populate a cone of
length 1.2\,kpc in the direction of NGC 2547 and assume 
reddening increases linearly at about $E(R_{\rm c}-I_{\rm
c})=0.1$\,kpc$^{-1}$. Stars are given intrinsic $R_{\rm c}-I_{\rm
c}$ using an empirical 5\,Gyr isochrone, generated as described
in section~\ref{isochrone}\footnote{The empirical 5\,Gyr isochrones
generated from the B02 and DM97 models are
so similar that it makes no difference which is used for the
simulation. The isochrones are also age-insensitive for ages $\geq
500$\,Myr.}.  $I_{\rm c}-Z$ colours are assigned using a polynomial
relationship between $R_{\rm c}-I_{\rm c}$ and $I_{\rm c}-Z$
found for {\em all} stars in our catalogue with $17<I_{\rm c}<19$.

We generate a simulated catalogue corresponding to 5 square
degrees of sky coverage -- about 6 times larger than our actual
survey area and sufficient to ensure that correction for
contamination plays little role in the {\it statistical} uncertainties in
the final LFs and MFs.  To account for measurement uncertainties and
completeness we have injected the simulated stars into our dataset
using the same ``ghosting'' technique described in
section~\ref{complete}. To restrict the numbers of ghosts to $<5$ per
cent of the number of stars in the genuine data we have clipped the
input catalogue to include only those stars with $12<I_{\rm c}<21.5$
and which lie sufficiently close to the NGC 2547 sequence to stand any
chance of contaminating the final membership catalogue.

After running the simulated data through the {\sc cluster} pipeline we
recover 7697 of the original 8391 ghosts, of which, 646 had bad quality
flags. Figure~\ref{contamfig} shows a subset of the recovered field
star ghosts, corresponding to 0.855 square degrees on the sky (we
simply plot a fraction of the recovered catalogue). From this the
reader can judge the level of contamination by comparison with
Fig.~\ref{cmdri}.  We pass the full 5 square degree simulated catalogue
through our membership selection procedure to find the subset of field
stars that would have been selected as candidate members. LFs/MFs
(corrected for incompleteness, see sections~\ref{lf} and~\ref{mf}) can
be obtained for this contaminating sample from whatever area of the
real catalogue we choose to construct the cluster LF/MF. This has the
additional advantage of accounting for differing sensitivities due to
the overlap regions or variations in exposure time and seeing.

Example ``observed'' (i.e. uncorrected for completeness) contaminating
field star LFs (taken from the full survey area), expressed in numbers
of stars per square degree are shown in Fig.~\ref{contamlf}. The main
feature of the LF is a broad peak at $I_{\rm c}\simeq16-17$, which is
apparent as an enhancement in the density of contaminating stars
in Fig.~\ref{cmdri}, and which is due to a maximum in the space density of
stars with absolute $I_{\rm c}\simeq 9$.  If the DM97
empirical cluster isochrone is used to select candidate members there
are an average 127 contaminating stars per square degree expected for the range
$14.25<I_{\rm c}<20.25$.  
The B02 empirical isochrone is fainter than that of
DM97 for $R_{\rm c}-I_{\rm c}>1.5$, resulting in more contamination at
these colours and a total of 179 contaminants per square degree for
$14.25<I_{\rm c}<20.25$.

A test of our contamination simulation is provided by fields 51 and 52,
which are 60-75 arcminutes from the cluster centre. There is
evidence (see section~\ref{masseg}) that the density of cluster
candidates is much lower in these fields than in the cluster centre, or
possibly zero. If this is the case, then the candidate members in these
fields represent the contamination level.

We find 11 candidate members with $14.25<I_{\rm c}<20.25$ in
fields 51 and 52 when selected using the DM97 isochrone or 13 when
using the B02 isochrone.  Our contamination simulations predict 
we should have seen 12.4$\pm0.5$ or $17.5\pm0.6$ field dwarf contaminants
respectively.  
This close agreement offers reasonable support for our contamination model,
although we must bear in mind two issues. First, the 8\,pc sample is
quite small: there are only 138 stars belonging to 103 systems defining
the field star LF. Second, the 8\,pc sample is probably complete to
$M_{I_{\rm c}}\simeq 12$, but thereafter is likely to be increasingly
incomplete (Reid, Gizis \& Hawley 2002). The field dwarfs that
contaminate our cluster sample will be those of the right colour and
distance to fall within the photometric selection bounds. More distant
intrinsically bright stars will be too blue, and very close
intrinsically faint stars will be too red. The contribution to the
contamination will be dominated by a $\sim 1$ mag range of the nearby
star LF, at a distance commensurate with their coincidence with the NGC
2547 PMS. The consequent uncertainties on each of the histogram points
in Fig.~\ref{contamlf} arising from Poisson errors in the nearby star
LF are therefore of order 20 per cent at $I_{\rm c}\simeq16$, about 30
per cent at $I_{\rm c}\simeq20$ and in all cases larger than the
statistical errors in our simulations.

More significantly, an unreddened dwarf with $M_{I_{\rm
c}}\simeq 12$ has $R_{\rm c}-I_{\rm c}\simeq 1.95$ and will contaminate
the PMS of NGC 2547 at $I_{\rm c}\simeq 18.5$ for the DAM97 isochrone,
or $I_{\rm c}\simeq 19.0$ for the B02 isochrone. At fainter
magnitudes in NGC 2547 it is probable that the contamination by field
dwarfs is {\em underestimated} by the simulation. A comparison of the
5.2\,pc sample which {\em is} believed to be complete beyond $M_{I_{\rm
c}}\simeq 12$ (but of course contains far fewer stars), suggests
that the 8\,pc LF could be underestimated by a factor of 2 for $11<M_{I_{\rm
c}}<14$ (Reid \& Hawley 2000). Insufficient area at large distances from
the cluster centre has been observed to empirically rule out this
possibility. For instance there are either 1 or 2 candidate members
with $18.75<I_{\rm c}<20.25$ lying in fields 51 and 52, depending on
whether the DAM97 or B02 isochrones are adopted. The contamination
simulation predicts we should have seen $0.80\pm0.13$ or $2.20\pm0.21$
respectively.

\section{Results}

\subsection{Mass segregation}
\label{masseg}

\begin{figure}
\vspace{195mm}
\includegraphics{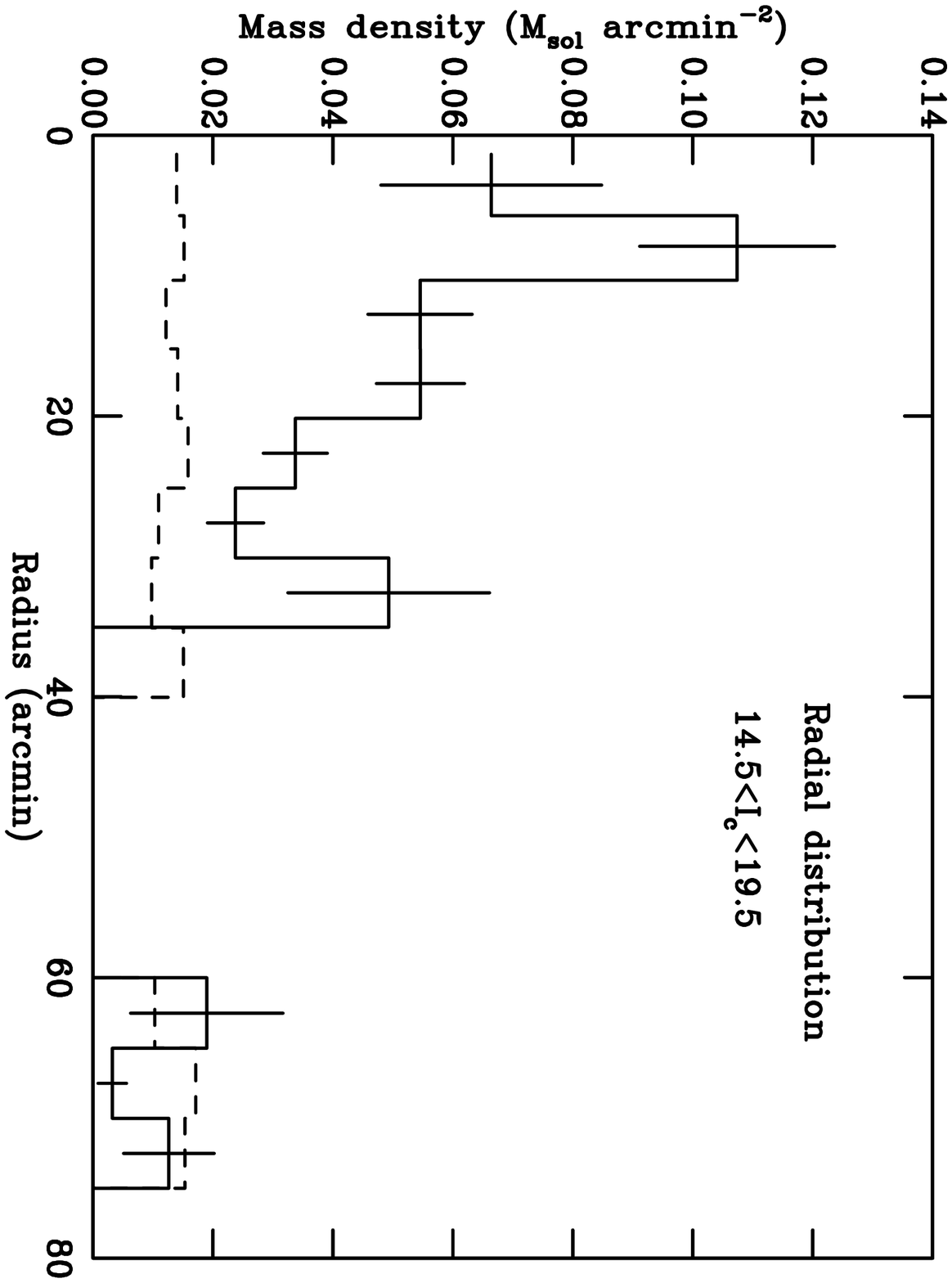}
\includegraphics{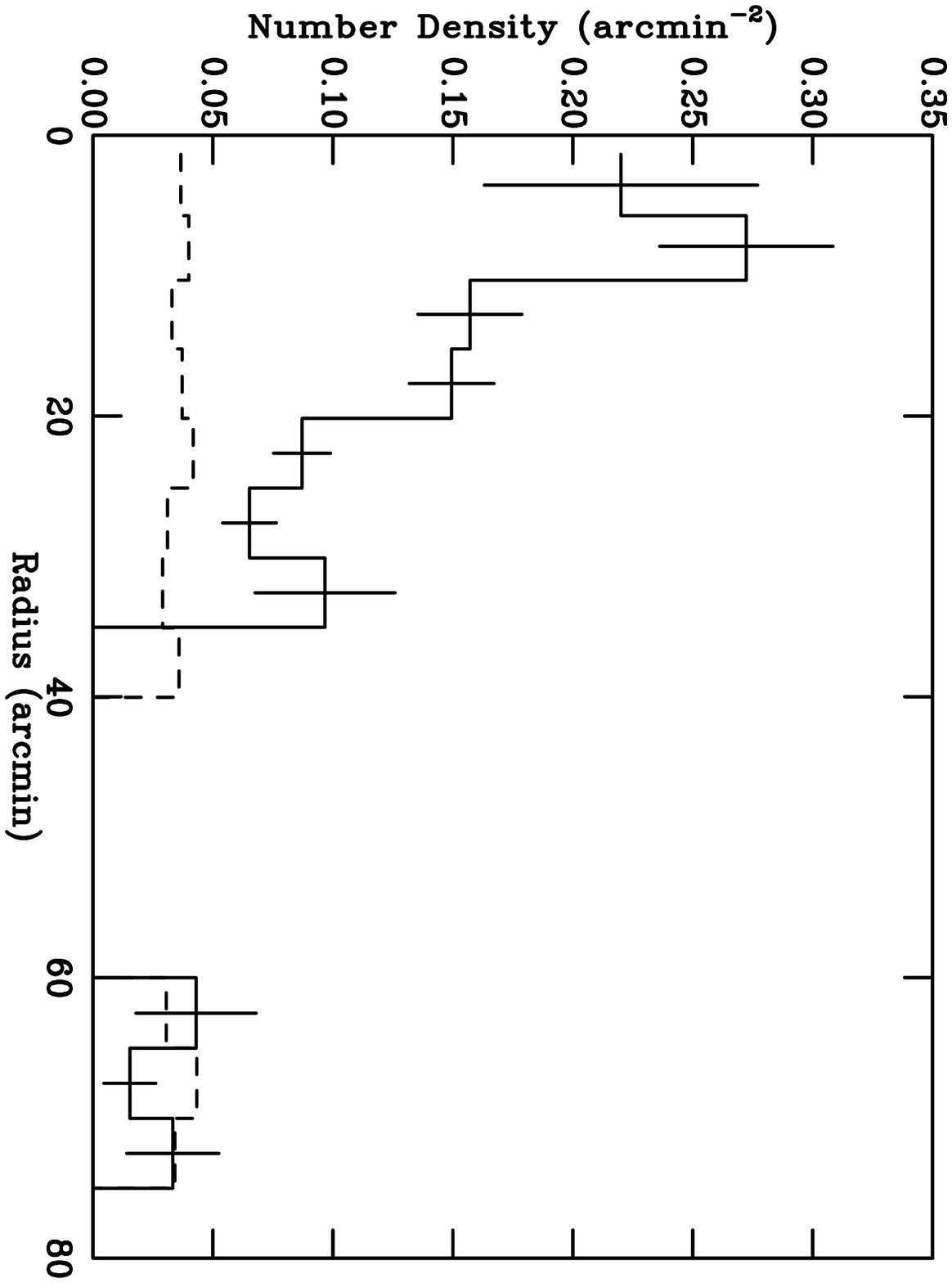}
\includegraphics{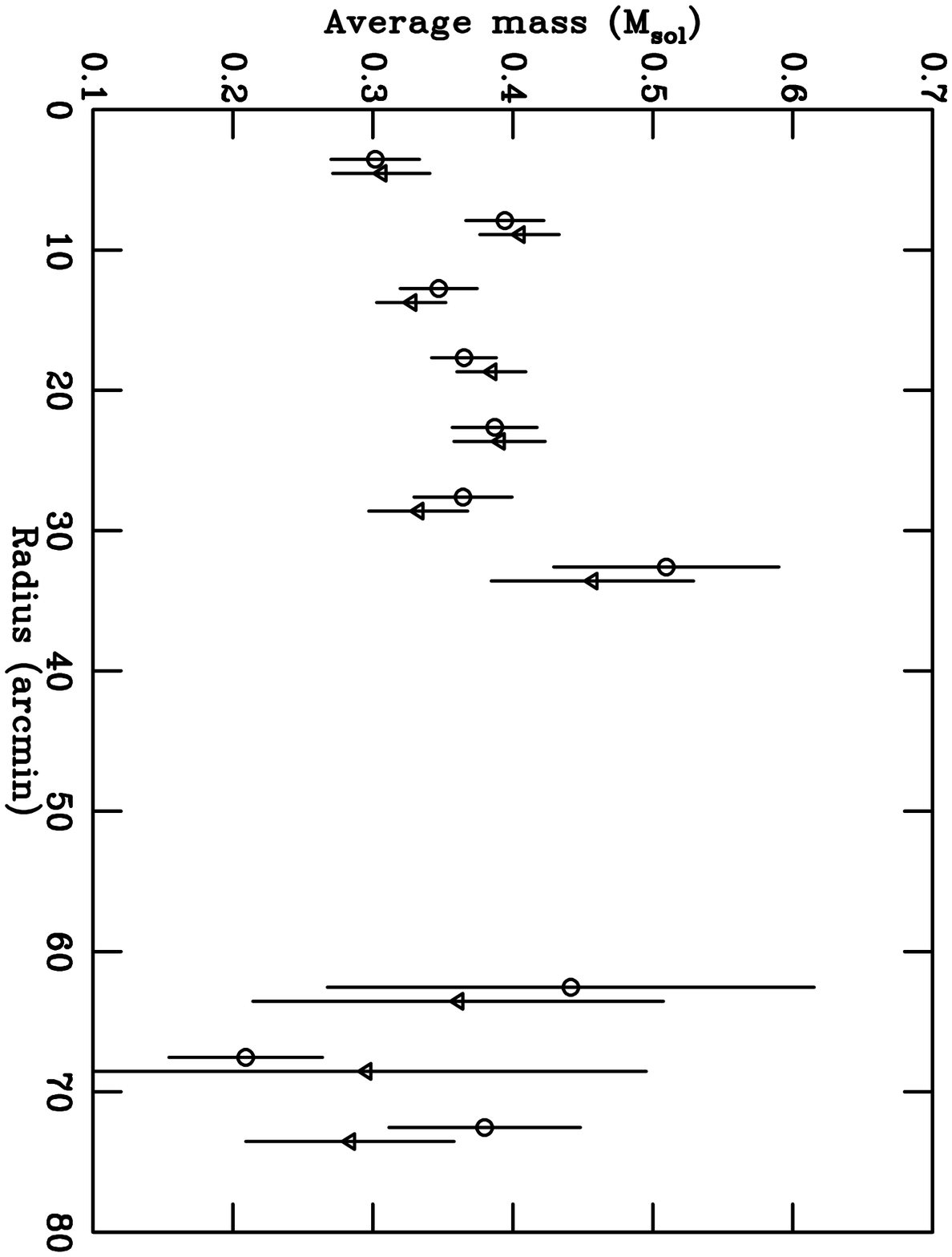}
\caption{(Top) The (completeness-corrected)
   spatial mass density of candidate members with $14.5<I_{\rm
    c}<19.5$ (based on the
  DAM97 isochrone) as a function of radius. The dashed
    line is the equivalent plot for candidate members selected 
    from our simulated field dwarf
    catalogue (which has much smaller statistical errors) scaled to
    the correct survey area. (Middle) The (completeness-corrected)
    spatial density of candidate member {\em systems}, the dashed line
    indicates the candidate member systems selected from our simulated
    field dwarf catalogue. (Bottom) The average mass per star
    system. Circles show the results from the DAM97
    isochrone, triangles show the equivalent analysis using the
    B02 isochrone. The B02 points have been offset to the
    right by 1 arcminute for clarity.}
\label{massdensity}
\end{figure}

\begin{table*}
\caption{The results of King model fitting to the radial number density
  profiles.}
\begin{tabular}{lcccccc}
& \multicolumn{6}{c}{{\bf D'Antona \& Mazzitelli 1997}} \\
Average Mass & 0.16$M_{\odot}$ & 0.16$M_{\odot}$ & 0.50$M_{\odot}$ & 0.50$M_{\odot}$ & 1.89$M_{\odot}$ & 3.85$M_{\odot}$ \\
\hline
&&&&&&\\
$r_{c}$ (arcmin)              & $16.5^{+5.6}_{-3.6}$ & $17.9^{+6.0}_{-4.7}$   &
$14.7^{+3.1}_{-2.5}$ & $15.4^{+3.9}_{-3.1}$ & $8.6^{+2.4}_{-1.9}$ &
$1.4^{+1.8}_{-1.3}$  \\
normalisation (arcmin$^{-2}$) & $0.141^{+0.029}_{-0.024}$ &
$0.147^{+0.028}_{-0.028}$ & $0.222^{+0.038}_{-0.033}$ &
$0.227^{+0.037}_{-0.032}$ & $0.195^{+0.057}_{-0.047}$ &
$0.289^{+2.248}_{-0.126}$  \\
background (arcmin$^{-2}$)    & 0.012 & $0.008^{+0.008}_{-0.007}$ &
0.024 & $0.021^{+0.007}_{-0.008}$ & 0.000 & 0.000 \\
$\chi^{2}/n_{\rm dof}$        & 3.7/6 &3.5/5 & 5.8/6 & 5.6/6 & 2.2/3 &
2.1/3  \\
$n(25)$ & 83.2 & 89.6 & 123.1 & 129.3 & 70.0 & 8.3  \\
$n(r_{t})$&138.6&153.9 & 195.9 & 209.2 &95.6 & 9.3  \\ 
&&&&&&\\
& \multicolumn{6}{c}{{\bf Baraffe et al. 2002}} \\
Average Mass & 0.16$M_{\odot}$ & 0.16$M_{\odot}$ & 0.51$M_{\odot}$ & 0.51$M_{\odot}$
 & & \\
\hline
&&&&&&\\
 $r_{c}$ (arcmin) & $11.3^{+4.2}_{-3.2}$ & $11.7^{+6.1}_{-4.4}$ &
$14.5^{+3.1}_{-2.4}$ & $16.2^{+3.9}_{-3.2}$ & & \\
normalisation (arcmin$^{-2}$) & $0.131^{+0.045}_{-0.035}$ &
$0.132^{+0.044}_{-0.038}$ & $0.208^{+0.035}_{-0.032}$ &
$0.220^{+0.034}_{-0.033}$ & & \\
background (arcmin$^{-2}$) & 0.0267 &
$0.026^{+0.009}_{-0.011}$ & 0.0257 & $0.019^{+0.008}_{-0.007}$ & & \\
$\chi^{2}/n_{\rm dof}$ & 2.0/7 & 1.9/6 & 9.9/6 & 9.1/5 & & \\
$n(25)$ & 60.6 & 62.8 & 114.5 & 128.4 & & \\
$n(r_{t})$& 88.7 &92.8 & 181.3 & 211.9 & &  \\ 
\end{tabular}
\label{kingtable}
\end{table*}

\begin{figure}
\vspace{215mm}
\includegraphics{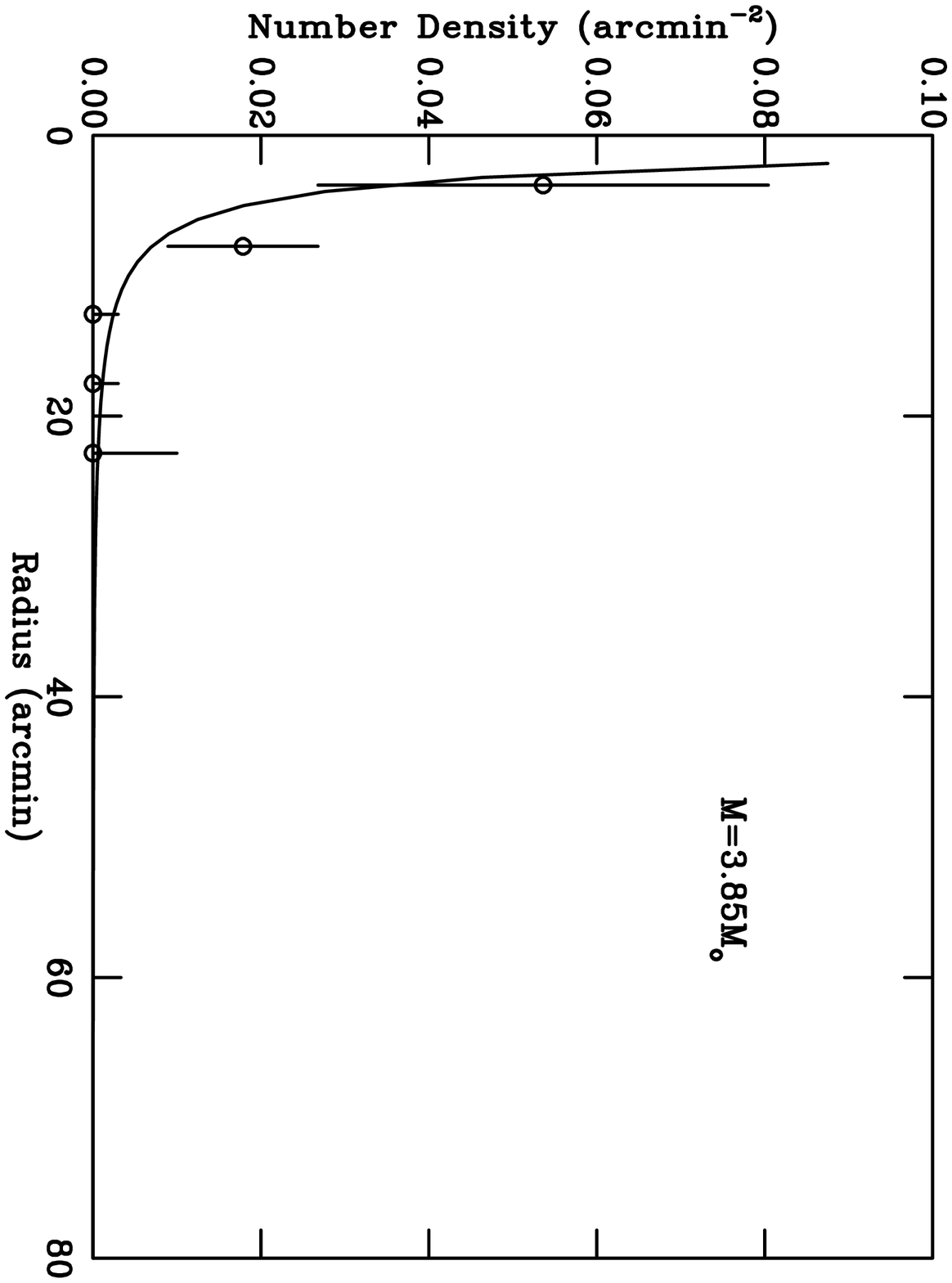}
\includegraphics{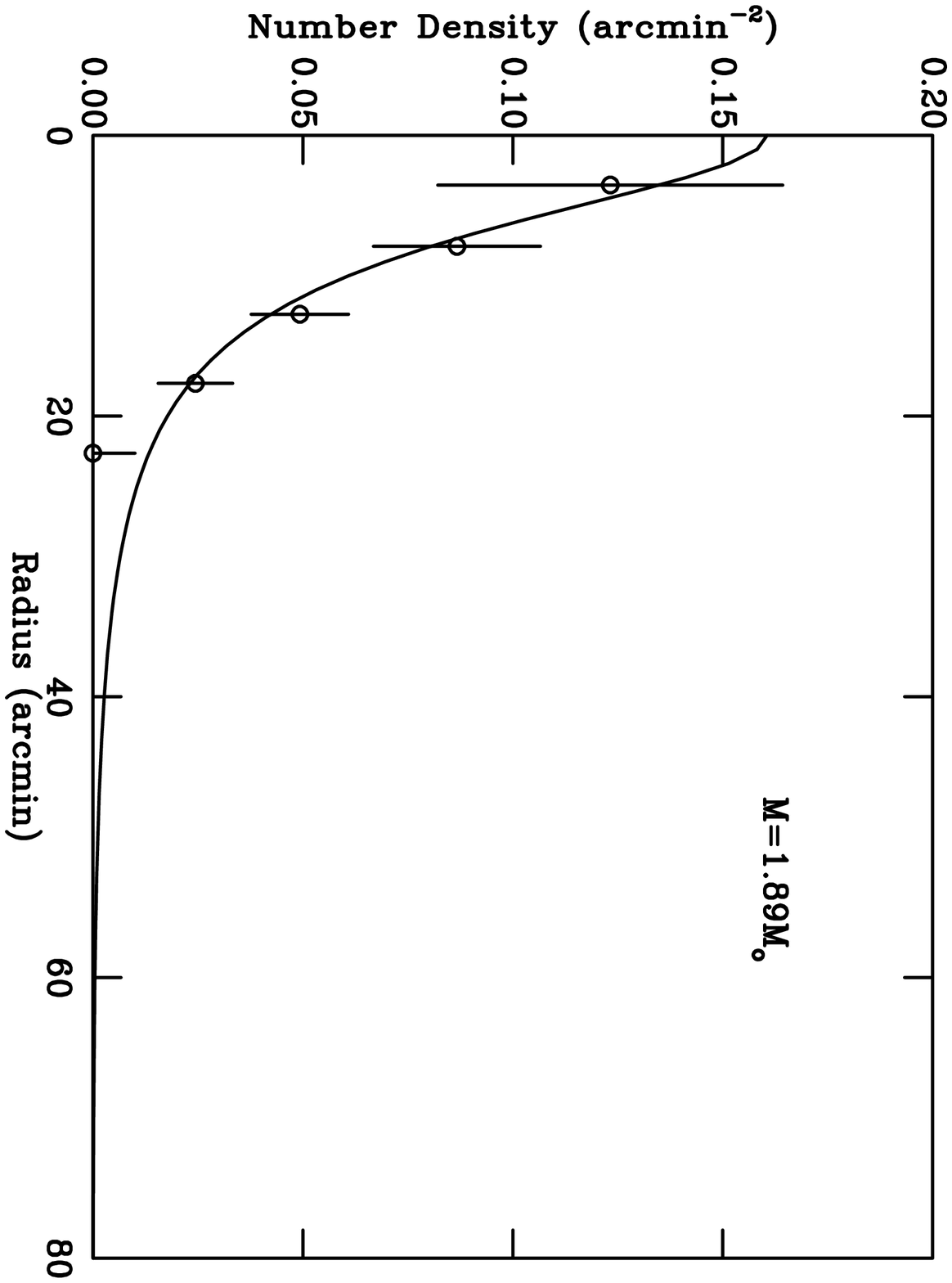}
\includegraphics{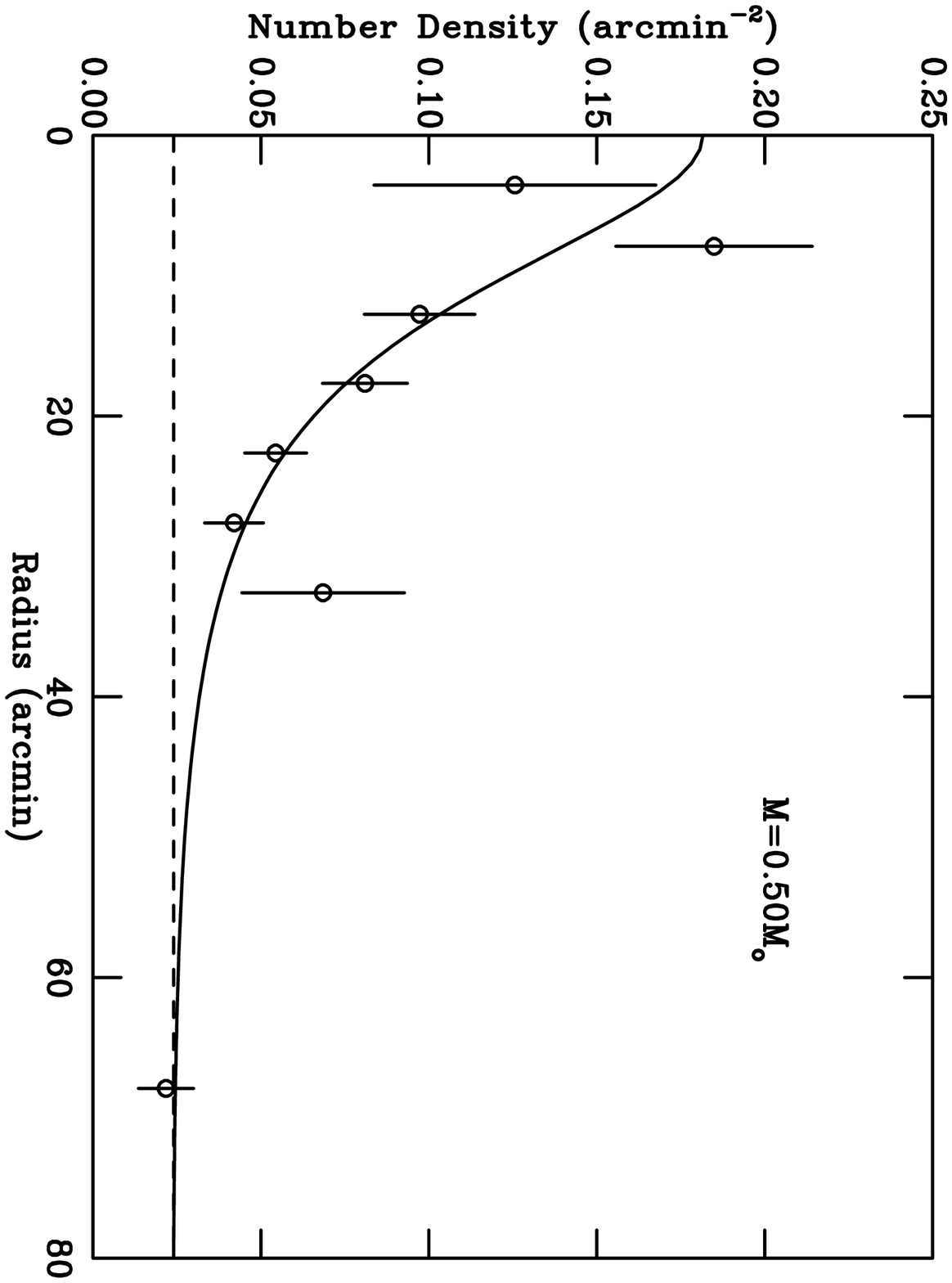}
\includegraphics{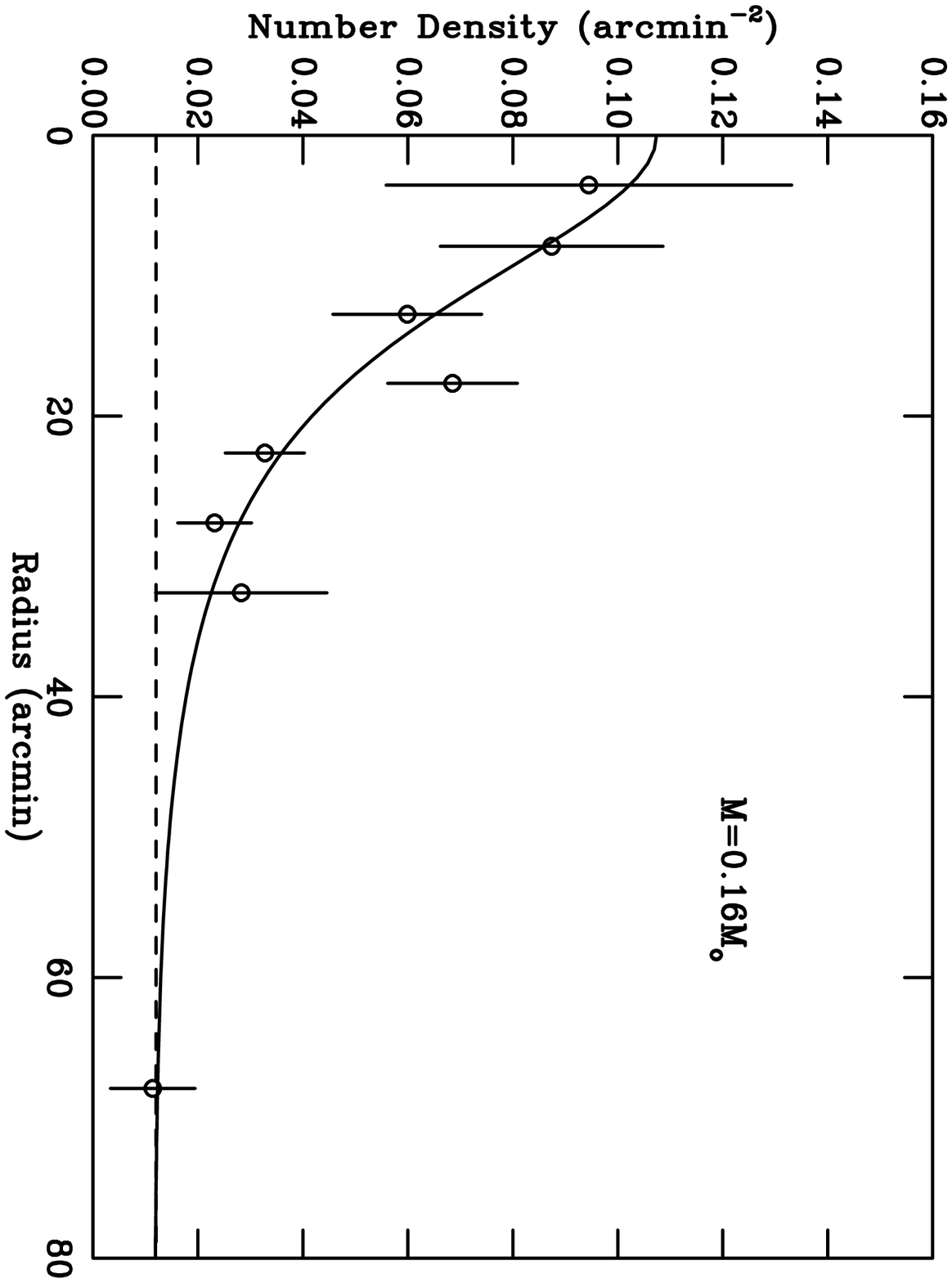}
\caption{Radial number density for systems of progressively lower
  average mass. System masses have been calculated according to the
  DAM97 isochrones. The solid lines indicate the King profile fits
  assuming a fixed background density of contaminating field systems,
  which is indicated by a dashed line. The higher mass fits are assumed
  to have zero contamination (see text).}
\label{kingplot}
\end{figure}

The work of Littlefair et al. (2003) showed that high mass stars
($>$3\,$M_{\odot}$) are more centrally concentrated in NGC 2547 than
their lower mass siblings. If this trend continued to even lower
masses then any derived low-mass LF or MF for NGC 2547 would
depend on the radius out to which candidate members are included,
unless this radius can be made large enough to encompass most cluster
members of all relevant masses.  However, the possible problems with
uncertain contamination lead to a conflicting need to minimise the
radius used for LF/MF estimation.

To look for mass segregation in NGC 2547 we need to assign a mass to
each of our candidate members. For candidates classified as single (see
section~\ref{select}) masses are assigned using a relationship between
$I_{\rm c}$ and mass that comes from the same empirical isochrone used
to select the members. A complication arises for possible binary
systems in that a binary may have twice the (system) mass of a single
star with the same colour. In this paper we assume that the $\sim30$
percent of candidate members classed as possible binary systems, on the
basis that they are $>0.5$ mag brighter than the single star isochrone,
are {\em all} unit mass ratio binary systems with equal luminosity
components. We subtract 0.75 mag from their $I_{\rm c}$ values,
calculate the mass of each star as if it were a single star and assume
that the system mass is double this. We are actually sensitive to
binaries with $q\geq0.35-0.65$ (depending on the colour -- see
Fig.~\ref{binfrac} and section~\ref{binary}) but, because binaries with
$q<1$ are redder than the intrinsic primary star colour, we find that
this approximation still yields a {\em system} mass that is within
about 10 per cent of that which would be determined using the correct
$q$ value. 

Figure~\ref{massdensity} shows the completeness-corrected
mass density (per square arcminute)
and number density (of systems) for candidate NGC 2547 members with
$14.5<I_{\rm c}<19.5$ (corresponding to $0.71>M>0.057M_{\odot}$ for the
DAM97 isochrone or $0.72>M>0.060M_{\odot}$ for
the B02 isochrone), selected using the DAM97
isochrone. The bottom panel of this figure shows the
average mass for the candidates.  The magnitude limits were chosen to
(a) avoid strong contamination by background giants and (b) avoid
any strong spatial
dependence in the completeness corrections (see section~\ref{complete}).  The
dashed histograms show the same quantities for the simulated field
dwarf contamination. The plots show the expected concentration of
mass towards the centre of the cluster, but there is no sign of any
decrease in the average mass with radius. This leads us to believe that
there is {\em no} strong mass segregation over this mass range. The
result is not model dependent. The bottom panel shows the radial
dependence of the average mass obtained by repeating the analysis using
candidate members and masses assigned using the B02 cluster
isochrone. However, the analysis is affected by increasing background
contamination as a function of radius and does not yield any {\em
  quantitative} limit on the amount of segregation that could be present.

To overcome these problems we divided the magnitude-selected 
sample into two subsets with masses above and below 0.25$M_{\odot}$,
corresponding to $I_{\rm c}\simeq 17.0$ for single stars with both sets of
model isochrones. Using the DAM97 isochrone, 
the average masses of {\em systems} in these two subsets are
0.16\,$M_{\odot}$ and 0.50\,$M_{\odot}$
respectively. The bottom two panels of Fig.~\ref{kingplot} show the radial number density
profiles for the subsets. The horizontal dashed lines on these plots
represent the average number density of field objects found for
similarly defined subsets in our simulated contamination catalogue.

We fitted these profiles with King functions (see King 1962) plus a
constant background density and fixed the
tidal radius of the cluster using $r_{t}=1.46\,M_{c}^{1/3}$ (Pinfield,
Hodgkin \& Jameson 1998; Jeffries et al. 2001), where
$M_{c}$ is the mass of the cluster. Assuming
$M_{c}=450M_{\odot}$ (see section~\ref{clustermass}) 
and a distance modulus of 8.1, then $r_{t}=92$ arcminutes. 
In one set of fits the background density was fixed at
the average value found from the simulated contamination catalogue and in
another the background was a free parameter. We were careful
to integrate the King profile over the 5 arcminute radial bins that
were necessary to accumulate sufficient counts for chi-squared fitting
to be valid. Because all the data lie well inside
the tidal radius, the best-fit core radii and normalisations of the King
functions are quite insensitive to $r_{t}$.
The results of our fitting are shown in
Table~\ref{kingtable}. For comparison we modelled the radial
number density profiles of two higher mass subsets of stars from the membership
catalogues of N02. The subsets are chosen with $V<9$
(2.8 to 5.9\,$M_{\odot}$, average mass 3.85$M_{\odot}$) and
$9<V<13$ (1.1 to 2.8\,$M_{\odot}$, average mass 1.89$M_{\odot}$).
Masses were assigned
to individual {\em systems} using the best-fit empirical $V,V-I_{\rm c}$ DAM97
isochrone described by N02. The N02 catalogue
does not extend beyond a 25 arcmin radius and so we could not 
fit the background density, but it is expected to be very small for
these brighter subsamples and
so we fixed it at zero (see Littlefair et al. 2003). 

The results in Table~\ref{kingtable} indicate a significant difference
in spatial distribution for the high and low-mass stars. 
Note that because there are only eight stars in the highest mass 
subsample we considered, the quoted
parameters and uncertainties (based on a chi-squared fitting technique)
should not be taken too literally. The mass segregation
implied by the much smaller core radius compared with the 
lower mass subsamples 
has however been confirmed, using a
number of statistical tests, by Littlefair et al. (2003). 
Our results show only marginal
evidence for any further mass segregation below 1\,$M_{\odot}$.
To put this on a more quantitative basis we fit
simultaneous King models to the two subsamples with average mass of
0.50\,$M_{\odot}$ and 0.16\,$M_{\odot}$ using the ratio of the core
radii, expressed as $r_{c, 0.16}/r_{c, 0.50}$, as a free parameter. We
find best-fit ratios of $1.13^{+0.40}_{-0.30}$ for the case of
fixed background values and the DAM97 isochrone, or $1.15^{+0.52}_{-0.37}$ if the background
densities are allowed to vary. The 95 per cent upper limits to the
ratio are 1.92 and 2.16 respectively.

We have also fitted a simple power law of the form $r_{c} \propto
M^{-\beta}$ to these two points plus the core radius of the subsample
with an average mass of 1.89\,$M_{\odot}$. We would expect $\beta=0$ if
no mass segregation is present over this mass range or
$\beta\simeq0.5$ in the case of dynamical equipartition (Pinfield et
al. 1998). We find $\beta=0.26\pm0.13$ or $\beta=0.30\pm0.15$ for the
cases of a fixed or free background respectively, and 90 percent
confidence intervals of $0.07<\beta<0.45$ and $0.09<\beta<0.51$.

Selecting candidate members and assigning masses using the B02
isochrone we obtain essentially indistinguishable results for the
subsample with average mass 0.51\,$M_{\odot}$, but in the lower mass
subsample, the empirical and model background estimates are
significantly higher. This results in smaller core radii estimates (see
Table~\ref{kingtable}), although consistent with those from the DAM97
modelling within the statistical uncertainties. The ratio $r_{c,
0.16}/r_{c, 0.51}= 0.78^{+0.34}_{-0.14}$ or $0.74^{+0.39}_{-0.16}$ for
the cases of fixed or freely fitted background densities. The 95 per
cent upper limits on the ratio are 1.48 and 1.57 respectively.

We conclude that there is evidence for strong mass segregation between
stars with $M>2.8\,M_{\odot}$ and lower mass stars; weaker evidence for
mass segregation of stars above and below $1$\,$M_{\odot}$; that for
lower mass stars ($<0.7\,M_{\odot}$) the radial distributions are
consistent with {\em no} mass segregation; that strong mass segregation
of the form $r_{c}\propto M^{-\beta}$ with $\beta>0.5$ for
$0.1<M<2.8\,M_{\odot}$ is ruled out at the 95 per cent level, and that
growth of the core radius by as much as a factor of 2 between
0.16\,$M_{\odot}$ and $0.50\,M_{\odot}$ is quite unlikely in the case
of the DAM97 modelling and ruled out for the B02 models.

\subsection{Luminosity function}
\label{lf}

\begin{figure}
\vspace{140mm}
\includegraphics{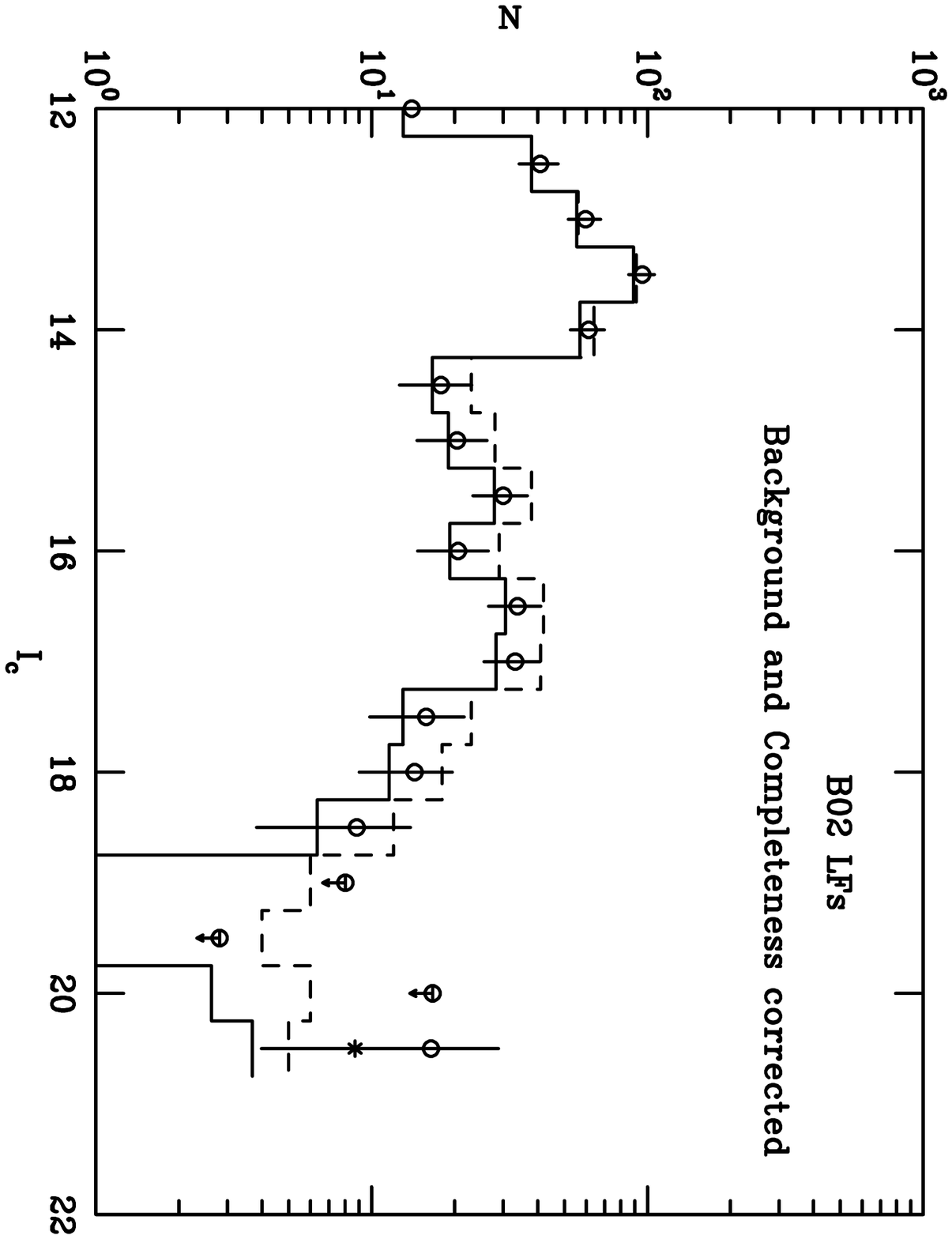}
\includegraphics{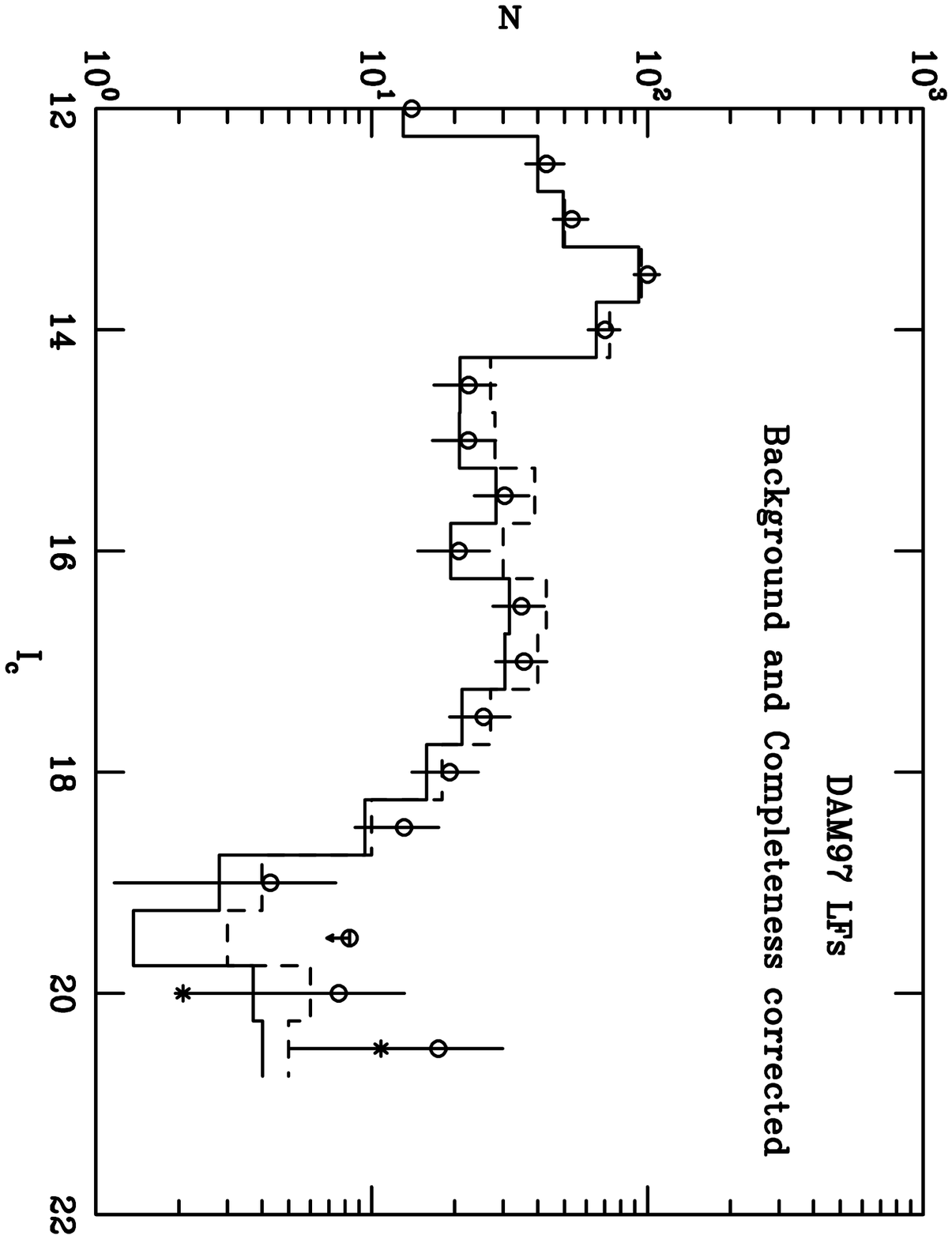}
\caption{(Top) The LF determined using the B02 models to assess cluster
  membership and contamination by background for the inner 25 arcminute
  radius of NGC 2547. The dashed line shows the LF of all stars
  satisfying the membership criteria. The solid line shows the LF after
  subtraction of the estimated contamination and the circles with
  errors bars show the contamination-subtracted LF after correction for
  incompleteness. (Bottom) Equivalent plots using the DAM97 models
  for membership selection. Points shown with downward pointing arrows
  are 2-sigma upper limits. The asterisks demonstrate the effect of
  doubling the background contamination at those magnitudes.
}
\label{lfplot}
\end{figure}
\begin{figure}
\vspace{70mm}
\includegraphics{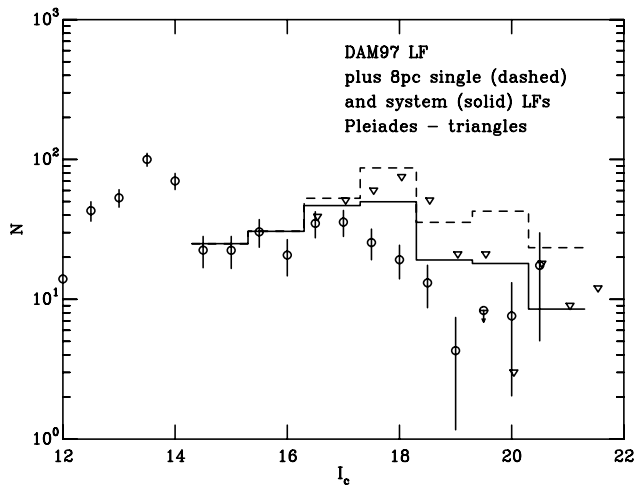}
\caption{The contamination-subtracted and completeness corrected LF for
  NGC 2547 compared with scaled LFs for stars within 8\,pc of the Sun
  (from Reid \& Hawley 2001) and the Pleiades (from Moraux et
  al. 2003). The dashed line shows the LF for all nearby stars, the
  solid line shows the LF when binary systems among the nearby stars
  are assumed to be unresolved (see text). The triangles show the
  Pleiades LF, and have uncertainties similar to those on the NGC 2547
  points.
}
\label{lfnearby}
\end{figure}

Having established limits on how much mass segregation may be present
in NGC 2547 we estimate the cluster LF by binning stars in $I_{\rm c}$
magnitude bins out to some limiting radius. We choose a radius of 25
arcmin, because this significantly exceeds the core radii for stars of
all masses making our LF insensitive to any mild mass
segregation, and because beyond this radius the 
contamination density becomes equal to or greater than the density of
cluster stars, compromising the precision of the LF.

The King profile fits in the last subsection allow us to estimate what
fraction of cluster members lie inside this radius by comparing the
integral of the King profile out to the tidal radius, with that out to a
radius of 25 arcminutes. These results (for the differing mass subsets)
are given in Table~\ref{kingtable}. Of course the estimates of the
numbers out to the tidal radius are reliant on the assumption
that a King profile is the correct form for the spatial density
at large radii.

The encouraging feature from the point of view of this paper is that,
at least for the two lowest mass subsets, 
the fraction of cluster members inside a 25 arcminute radius only varies
in a narrow range between 58 and 68 percent depending on 
mass, background estimate and selected evolutionary model. This is a
consequence of the lack of demonstrable mass segregation and the relatively
consistent estimates of core radii from both the DAM97 and B02 models.
Additionally, although the background contamination density varies by
a factor of 2-3 between the DAM97 and B02 models in the lowest mass
subset, it does not dominate in either until radii greater than 25
arcminutes. As a result, when estimating the MFs and the LFs for stars
from the survey described in {\it this} paper, there is no
compelling reason to apply any differential correction due to 
mass segregation, although such an effect may be present at levels of
10 percent for $0.1<M<0.7\,M_{\odot}$.

To construct an LF we place cluster candidates into 0.5 mag bins
(in $I_{\rm c}$) and apply completeness corrections by incrementing
the binned value by the reciprocal of the completeness correction per
star. We do the same for the simulated sample of contaminating field
dwarfs selected as candidate cluster members and subtract this
from the cluster candidate LF, propagating the statistical uncertainties
according to the binomial distribution. The results are shown in
Fig.~\ref{lfplot} for the cases where the DAM97 or B02 isochrones were
used for membership selection. The effects of completeness correction
and the contamination subtraction are indicated.

The large peak in the LF at $12<I_{\rm c}<14.5$ is attributable to
contaminating background giants. Our contamination model does not
include these, so they are not subtracted correctly. At fainter
magnitudes the LF exhibits a shallow rise, reaching a peak at $I_{\rm
c}\simeq 17$ and then declines sharply, reaching a minimum at $I_{\rm
c}\simeq 19.5$ that is consistent with there being {\em no} objects in
NGC 2547 at this magnitude.
 Comparing the dashed lines, solid lines and points with
error bars in Fig.~\ref{lfplot}, we can see that the accuracy of the
contamination subtraction and completeness correction are unlikely to
have a great bearing on the above statements and there also seems to be
little model dependency in the results.

The LF appears to rise again for $I_{\rm c}>20$, but here we have
to caution the reader that the completeness corrections are large and
the contamination model may itself be incomplete at
colours corresponding to these magnitudes (see section~\ref{contaminate}). Our
estimates of the LF are probably upper limits at this
point, but Fig.~\ref{lfplot} shows that, even if the subtracted
contamination in the final three bins is increased by a factor of two,
it makes little difference to the overall LF shape.

The LF of NGC 2547 (which has a distance modulus of about 8.1) can be
compared with LFs found for other young clusters and the field. For
example, Barrado y Navascu\'{e}s et al. (2002) find that the $I$-band
LF for low mass members of the 70-90\,Myr old Alpha Per cluster peaks
at $M_{I}\simeq 10$ and then declines by a factor of three by
$M_{I}\simeq13$.  Moraux et al. (2003)
find a similar peak in the Pleiades. We have obtained the Pleiades data
from Moraux et al. (2003), corrected it for the 30 per cent
contamination in the brown dwarf regime suggested in that paper, found
the LF, shifted it to match the distance modulus of NGC 2547 and
applied an arbitrary normalisation to provide an approximate match to
NGC 2547 at $I_{\rm c}\simeq 17$. The comparison can be seen in
Fig.~\ref{lfnearby}, where statistical uncertainties on the Pleiades points are
quite similar to those in NGC 2547. The decline in the NGC 2547 LF appears to
be steeper than in the Pleiades (and Alpha Per) and the LF peak occurs
at $M_{I}\simeq9$, more like that found for the very young
($\sim5$\,Myr) Sigma Orionis cluster, by B\'{e}jar et al. (2001).  Of
course because the luminosity-mass relation will be a function of age
as young low-mass stars descend their PMS tracks, this is not
necessarily surprising.  B\'{e}jar et al. also show a (marginal)
minimum in the $I$ band LF for $11<M_{I}<13$, but their photometrically
selected sample may (like ours) be contaminated with non-members at
fainter magnitudes.
 
In Fig.~\ref{lfnearby} the NGC 2547 LF is compared with the LF of the
local 8\,pc sample from Reid \& Hawley (2000), which has been shifted
to the distance modulus of NGC 2547 and arbitrarily normalised to agree
with the NGC 2547 LF at $I\sim 15$. Here the comparison is far from
favourable, there appear to be significantly more faint stars in the
nearby sample.  Partly this must be an age effect -- when comparing
stars at similar absolute $I_{\rm c}$ magnitude in the field and NGC
2547 we are comparing stars with very different masses. For instance,
$18.5<I_{\rm c}<19.5$ corresponds to $0.095>M>0.057\,M_{\odot}$
according to the DAM97 NGC 2547 isochrone. Allowing for the distance
modulus of NGC 2547, the same magnitude range corresponds to
$0.197>M>0.124\,M_{\odot}$ in a group of stars with age 5\,Gyr. If the
mass function declines with mass over this range, then it would be
unsurprising to find more faint stars in the field sample. The local LF
also contains stars from many resolved binary systems (see
section~\ref{contaminate}). The majority of these would {\em not} be
resolved in NGC 2547 (or other young clusters) and so a straightforward
comparison with the nearby LF is not likely to be valid, as many faint
stars in NGC 2547 will be hidden in binary systems.  Instead we show in
Fig.~\ref{lfnearby} what the local LF would look like if none of the
binary systems were resolved. This accounts for some of the
discrepancy, but much remains -- and the discrepancy will be enhanced
at fainter magnitudes if the 8\,pc sample is incomplete.

\subsection{Mass function}
\label{mf}

\begin{figure}
\vspace{140mm}
\includegraphics{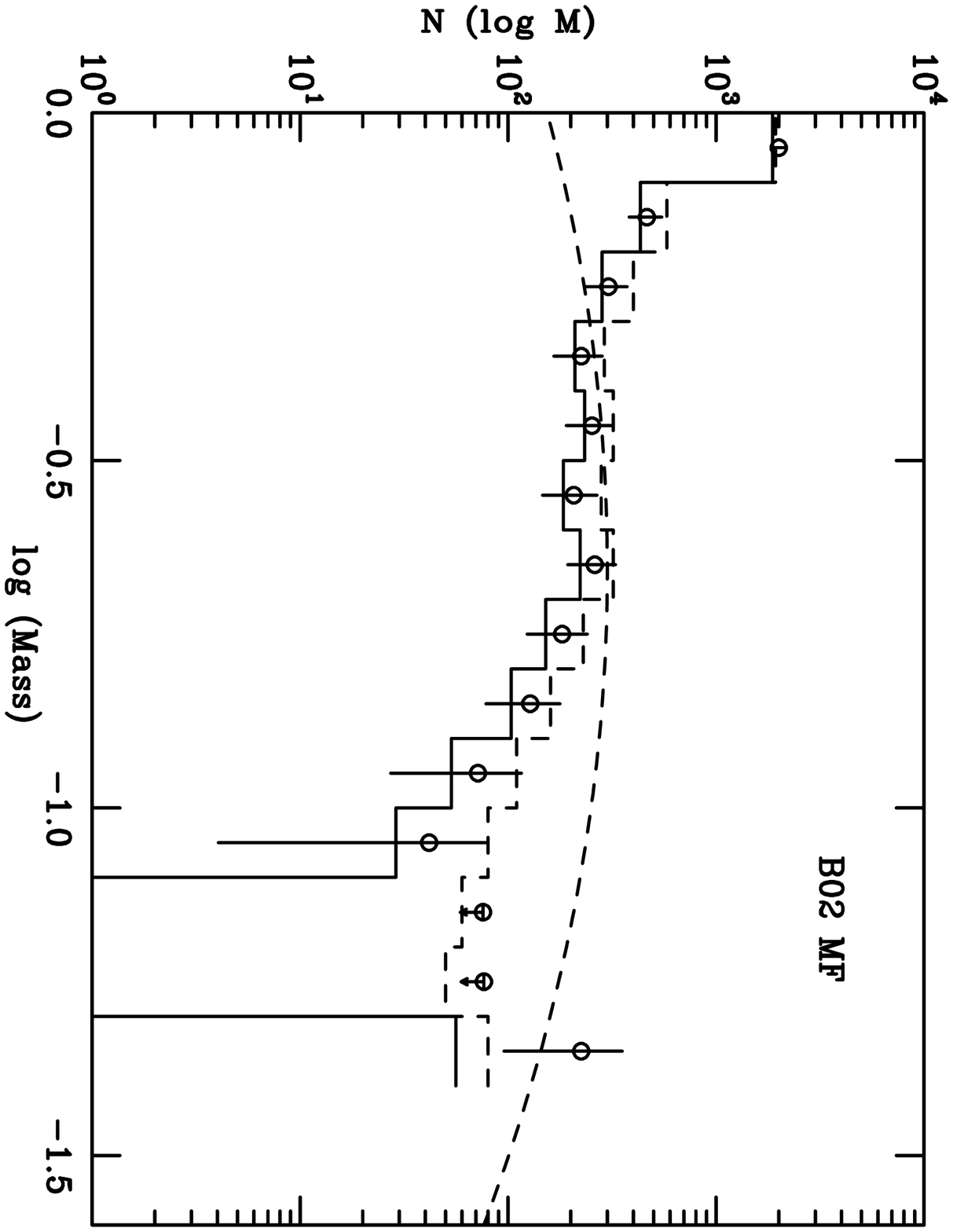}
\includegraphics{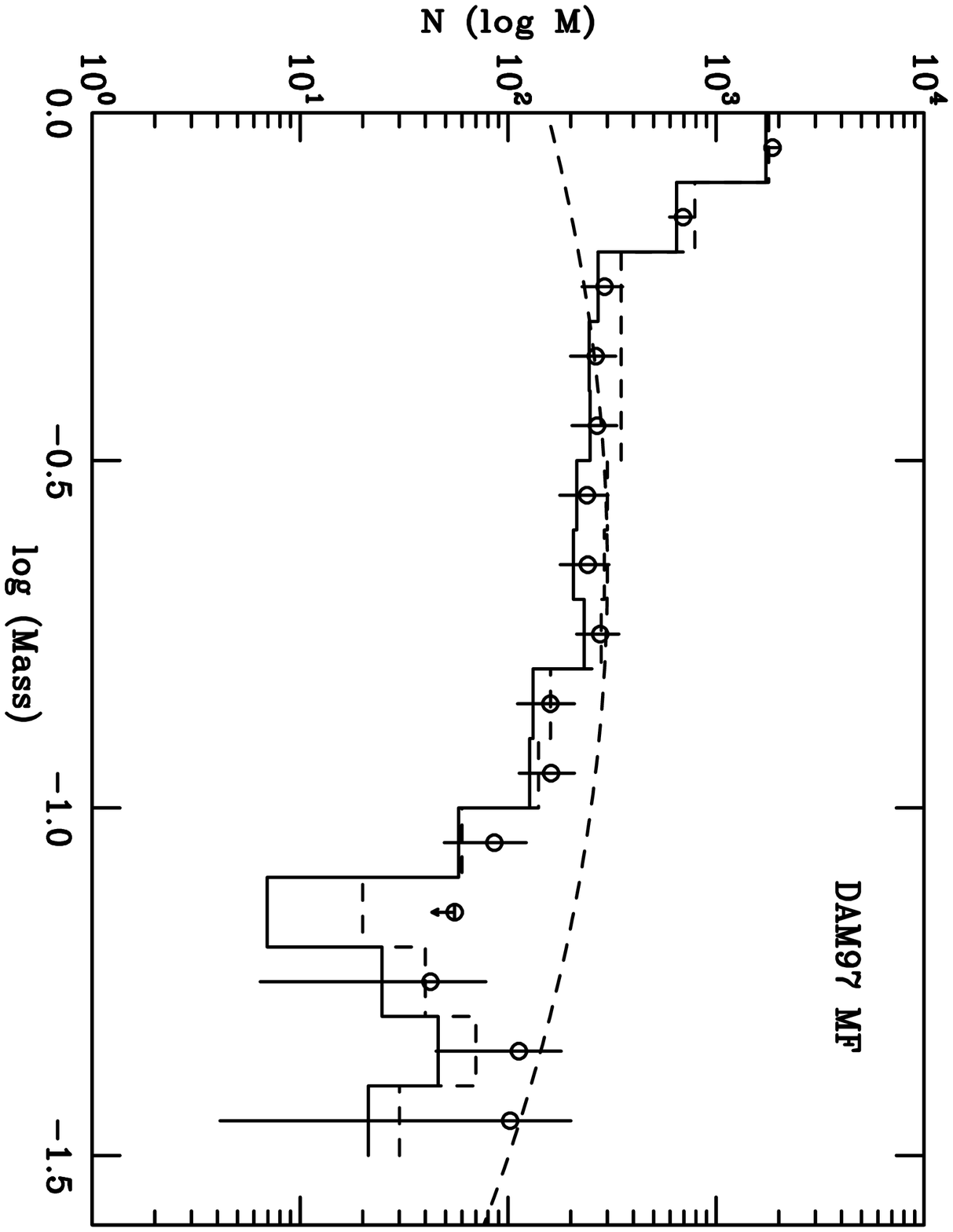}
\caption{(Top) The MF (determined using the B02 models)
  for the inner 25 arcminute
  radius of NGC 2547. The dashed histogram shows the MF of all stars
  satisfying the membership criteria. The solid histogram shows the MF after
  subtraction of the estimated contamination and the circles with
  errors bars show the contamination-subtracted MF after correction for
  completeness. (Bottom) Equivalent plots using the DAM97 models.
  Points shown with downward pointing arrows are 2-sigma upper limits.
The smooth dashed curves show the log-normal {\it system} IMF for 
the young disc and open clusters proposed by Chabrier (2003), with an
  arbitrary normalisation.
}
\label{mfplot}
\end{figure}

\begin{figure}
\vspace{70mm}
\includegraphics{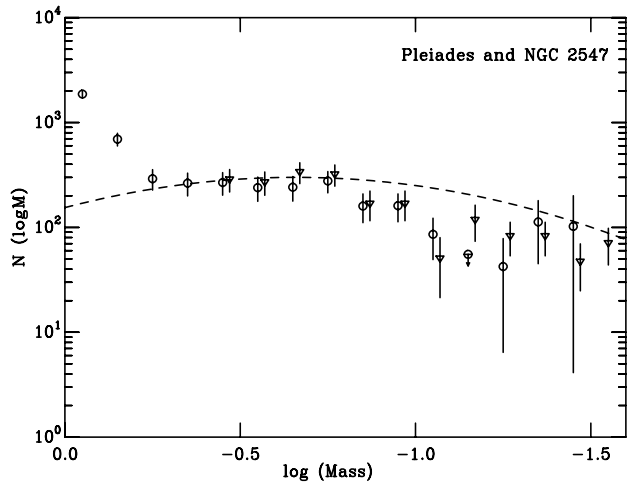}
\caption{(Top) The corrected MF 
  (for the inner 25 arcminute
  radius) of NGC 2547 (circles) compared with the MF of the Pleiades (triangles)
  derived from data presented by Moraux et al. (2003). Both MFs were
  calculated using empirical $I_{\rm c}$ versus mass relationships
  derived from the DAM97 models. An arbitrary normalisation has been
  applied to the Pleiades points as well as a small shift in $\log M$
  to clearly distinguish the two sets of points.
}
\label{pleiadesmf}
\end{figure}

The cluster MF is estimated by calculating the mass of
each candidate member using a mass-$I_{\rm c}$ relationship derived from the
isochrones that were used to fit the cluster CMDs. In this paper we
define the MF using the form $dN/d\log{M} \propto M^{-\alpha}$, 
where the canonical IMF for field stars derived by
\cite{salpeter55} would have $\alpha=+1.35$.  

Unresolved binarity is a problem in two ways: A
binary star that appears above the single star cluster locus will be
treated as one star with a slightly higher mass than either of its two
components. The second problem is that binary stars with unequal mass
components will have a derived primary star mass that is
approximately correct, but the hidden, lower mass secondary component
will not be included in the MF. Sagar (1991), Kroupa (2001)
and Chabrier (2003) show that these two effects could result in systematic
underestimates of $\alpha$ by between 0.3 and 1, depending on the mass
range considered, the binary fraction and the steepness of the MF.

In section~\ref{masseg} we discussed a simple correction for binarity
that results in reasonably accurate estimates of {\em system} masses,
but does not really assist us in generating the stellar MF because $q$
can only be loosely constrained for the photometric binaries and an
unknown fraction of the remaining ``single'' stars are actually low-$q$
binary systems.  Attempting to correct the observed MF for binarity is
important in estimating the total cluster mass, but not for comparing
cluster MFs with each other or with the field MF, so long as
corrections are applied consistently, or not at all. A caveat
is that there must be no significant differences in binary fraction or
$q$ distribution. There is however little evidence for any enhanced binary
frequency in younger clusters compared to the field (see Bouvier et
al. 2001 and section~\ref{binary}). 
Halbwachs et al. (2003) also conclude that both the binary
frequency and the $q$ distributions of field F-K binaries and their
counterparts in the Pleiades and Praesepe are indistinguishable.

We derive the MF of NGC 2547 in a similar way to that done for the
LFs. Each star is assigned to a bin (we use bins of size 0.1\,dex in
$\log$ mass), a completeness correction is applied and the MF due
to (completeness-corrected) contamination is subtracted. The MFs for
the cases where the DAM97 or B02 models were used are shown in
Fig.~\ref{mfplot} in the form of $dN/d \log M$ versus $\log M$.

The features apparent in the LFs are mirrored here. There
is a strong peak in the MFs for $0.7<M<1.0\,M_{\odot}$ which we
attribute to contamination by giant stars -- the {\em two}
highest mass points in Fig.~\ref{mfplot} should be considered extremely
conservative upper limits. Both plots show a
roughly flat MF for
$0.2<M<0.7\,M_{\odot}$, which then turns downwards towards the brown
dwarf regime. There is a minimum at $I_{\rm c}=19.5$ corresponding 
to a $\log M=-1.24$  
for the DAM97 isochrones and $\log M=-1.22$ for the B02 isochrones.
The MFs are consistent with zero at this point, unless the level of
background contamination has been seriously {\em overestimated}.
However, incompleteness in the local LF used to generate
the contaminating sample means that the converse is more likely,
especially for the the lowest mass two and three points of
the B02 and DAM97 MFs respectively. These should probably be considered
(increasingly conservative) upper limits to
the MF of NGC 2547. 

Many authors have described the MF of the field and young clusters in
terms of a power law ( $dN/d\log{M} \propto M^{-\alpha}$). There are
problems with this approach, not least the arbitrary way in which the
mass range for the fits is chosen (see below). For the range $-1.3<\log
M < -0.4$ we obtain $\alpha = -1.06\pm0.20$ for the DAM97 MF, with a
relatively poor chi-squared of 12.0 for 7 degrees of
freedom. Alternatively, we obtain $\alpha=-1.23\pm0.25$ for the B02
MFs, with a reduced chi-squared of 6.2.
However, it would also be perfectly reasonable to model both MFs with a
flat function ($\alpha \simeq 0$) for $-0.8< \log M < -0.2$ followed by
a rather steeper decline with $\alpha \simeq -2$ below this.

A value of $\alpha\simeq 0$ for $-0.8<\log M<-0.2$ is rather
typical of that found in the field and for a number of young clusters
such as the Pleiades (see Meusinger et al. 1996; Gould 1997; Kroupa
2001). It is also consistent with the results obtained for stars
overlapping this mass range in the NGC 2547 survey of N02. 
Chabrier (2003) hypothesises that the disc MF
has a log-normal form (shown as a smooth dashed curve with
arbitrary normalisation in Fig.~\ref{mfplot}), with a characteristic
mass (when binary systems are not resolved) of 0.22\,$M_{\odot}$ and
Gaussian sigma of 0.57\,dex. The field MF is not well constrained
below 0.1\,$M_{\odot}$, but Chabrier claims that an extension of
this log-normal form does a reasonable job of describing the very low
mass and substellar MF in a number of young clusters, notably the
Pleiades and Alpha Per. The authors that obtained the data upon which
this conclusion is based have tended rather to fit power law MFs
with indices of $-0.2<\alpha<-0.6$ over a variety of mass ranges within
the limits $-0.4<\log M <-1.6$ (e.g. B\'{e}jar et al. 2001; Barrado y
Navascu\'{e}s et al. 2002; Jameson et al. 2002; Moraux et al. 2003).

Our MF for NGC 2547 is reasonably consistent with Chabrier's log-normal
model down to masses of $0.2\,M_{\odot}$, but then declines more steeply.
When expressed as a power law, the MF for NGC 2547 also seems to drop
more rapidly towards lower masses than in most young clusters.
Much of this discrepancy may be down to the variation in mass limits over which
power-law fits are applied. A better way to compare clusters is to
derive MFs from the original data using the same magnitude-mass relationships.
Taking the $I_{\rm c}$ magnitudes for stars and brown dwarf candidates
in the Pleiades (Moraux et al. 2003), we have used the $I_{\rm c}$-mass
relationships used in this paper to calculate the Pleiades MF. As
discussed by Moraux et al., we apply no correction for mass segregation
and correct for 30 per cent contamination among the brown dwarf
candidates.

Figure~\ref{pleiadesmf} shows the comparison between the MFs of NGC
2547 and the Pleiades (derived using DAM97 models, an age of 120\,Myr
and a distance modulus of 5.6).
The MFs have been scaled to roughly agree with that of
NGC 2547 at $0.3\,M_{\odot}$. The agreement between the MFs is
exceptionally good down to the substellar limit, with a similar level
of statistical precision in both clusters. There is then a
suggestion that there are fewer brown dwarfs in NGC 2547 than the
Pleiades. This statement is based on one well defined
point in the MF (at $\log M=-1.15$), which is about 5-sigma discrepant
from the Pleiades point, and the knowledge that lower mass points
in NGC 2547 may well be afflicted with a higher level of contamination
than we have assumed. The Pleiades MF also features a $\sim 2\sigma$ dip 
at $\log M=-1.05$. Dobbie et al. (2002) have suggested that sharp
dips in the MF close to the substellar limit may be as a result of an
imperfect understanding of cool atmospheres and hence the
mass-magnitude relationship (see section 5.3).
The Pleiades MF {\em also}
seems a little inconsistent with the log-normal form proposed by
Chabrier (2003) and this is not changed if the MF is calculated using the
B02 isochrones. The Pleiades MF also features a $\sim 2\sigma$ dip 
at $\log M=-1.05$

Both datasets favour a narrower Gaussian sigma of around
0.4 in $\log M$, although the point at $\log M=-1.15$ in NGC 2547
would still lie well below even this model. 
Finally, as a cautionary tale, we find that a power law MF fitted
between $-1.6<\log M < -0.4$ in the Pleiades 
yields $\alpha = -0.77\pm0.12$ (and similar using the B02 models), 
but a fit between $-1.2<\log M <-0.7$ yields $-1.9\pm0.6$!

\subsection{The overall mass function of NGC 2547}

\begin{figure}
\vspace{70mm}
\includegraphics{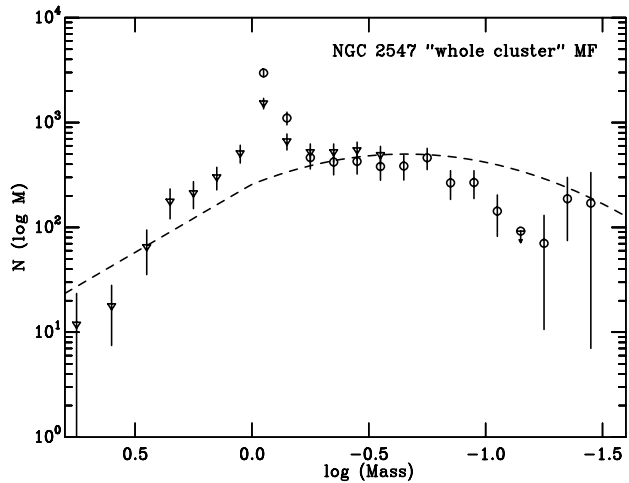}
\caption{The ``whole-cluster'' MF of NGC 2547 taken from the N02 survey (triangles) and
  this paper (circles) and using a 30\,Myr DAM97 isochrone. 
  Corrections have been made for completeness,
  and contamination (only for the data from this paper). Approximate
  corrections have been applied to include all objects out to the
  cluster tidal radius, accounting for the differing survey areas and
  mass-segregation (see text). The dashed line shows the form of the
  IMF proposed by Chabrier 2003).
}
\label{totalmf}
\end{figure}

\begin{table}
\label{mftab}
\caption{The ``whole cluster'' mass function for NGC 2547.}
\begin{tabular}{ccl}
$\log M$ & $N (\log M)$ & Notes \\ 
&&\\
0.75     & $12\pm 12$ & 1\\
0.60     & $18\pm 10$ & 1\\
0.45     & $65\pm 29$ & 1\\
0.35     & $177\pm 56$ & 1\\
0.25     & $212\pm 61$ & 1\\
0.15     & $301\pm 73$ & 1\\
0.05     & $508\pm 98$ & 1, 3\\
-0.05    & $1524\pm 175$& 1, 3\\
-0.15    & $662\pm 115$ & 1, 3\\
-0.25    & $463\pm 102$ & 2 \\
-0.35    & $420\pm 103$ & 2 \\
-0.45    & $426\pm 104$ & 2 \\
-0.55    & $381\pm 99$ & 2 \\
-0.65    & $384\pm 102$ & 2 \\
-0.75    & $461\pm 106$ & 2 \\
-0.85    & $266\pm 81$ & 2 \\
-0.95    & $267\pm 79$ & 2 \\
-1.05    & $143\pm 61$ & 2 \\
-1.15    & $<92$       & 2, 4 \\
-1.25    & $70\pm 60$  & 2, 5 \\
-1.35    & $187\pm 113$ & 2, 5\\
-1.45    & $170\pm 163$ & 2, 5\\
&&\\
\multicolumn{3}{l}{1 Data from N02}\\
\multicolumn{3}{l}{2 Data from this paper}\\
\multicolumn{3}{l}{3 Contaminated by giants}\\
\multicolumn{3}{l}{4 2-sigma upper limit}\\
\multicolumn{3}{l}{5 Probable additional contamination}\\
\end{tabular}
\end{table}

We can combine the low-mass MF
determined here with the cluster candidates from N02 to estimate a
cluster MF over two orders of magnitude in mass --
$0.06<M<6.0\,M_{\odot}$. We calculate an MF from the $V$ magnitudes of
the N02 survey by using a $V$-mass relationship defined by the
empirical 30\,Myr DAM97 isochrone used to select members in that paper.
For stars with $M>3\,M_{\odot}$ the isochrones of Schaller et
al. (1992) are used. The data are corrected for completeness, but 
no correction is made for binarity and there is assumed to be no
contamination.

To compare the MFs from the N02 and our survey we must account for the
differing survey areas and the (mass-dependent) spatial distribution of
the cluster candidates. The MFs are normalised to what would be seen out
to the tidal radius if King profiles accurately 
represent the spatial distributions. The corrections are applied in a
mass-dependent way using the four models (with fixed background) from
Table~\ref{kingtable}. For the record we multiplied the MF contribution
from $V<9$ stars by 1.16, the $9<V<13$ stars by 1.73, the $V>13$ stars
by 1.94 in the N02 data, and the $14.5<I_{\rm c}<17.0$ and $I_{\rm
  c}>17.0$ stars from this paper by 1.59 and 1.67 respectively.

The final MFs are plotted for both datasets in
Fig.~\ref{totalmf}. Where the datasets overlap in mass, the agreement
is reasonable. The N02 MF is higher for $M<0.8\,M_{\odot}$, but this is
probably because no contamination has been allowed for -- the MF from
this paper is far more reliable in this mass range. The ``spike'' in
the MF at $\simeq 1\,M_{\odot}$, caused by background giants is lower
in the N02 dataset because many giants were cleaned from
this sample by selection in the $B-V$ versus $V-I_{\rm c}$ diagram.
The cluster is not well fitted by the log-normal plus power-law
form suggested bt Chabrier (2003), but could be well represented over
the range $0.06<M<6.0\,M_{\odot}$ with three power laws, as suggested by
Kroupa (2001), with $\alpha\simeq 2.2\pm0.3$ for $1<M<6\,M_{\odot}$,
$\alpha \simeq -0.1\pm0.3$ for $0.2<M<0.7\,M_{\odot}$
and $\alpha \simeq -2.2 \pm 0.5$ for $0.06<M<0.2\,M_{\odot}$.
As an aid to future work, Table~\ref{mftab} list our best estimate of
the ``whole cluster'' MF for NGC 2547.

\subsection{Binarity}

\label{binary}

\begin{figure}
\vspace{135mm}
\includegraphics{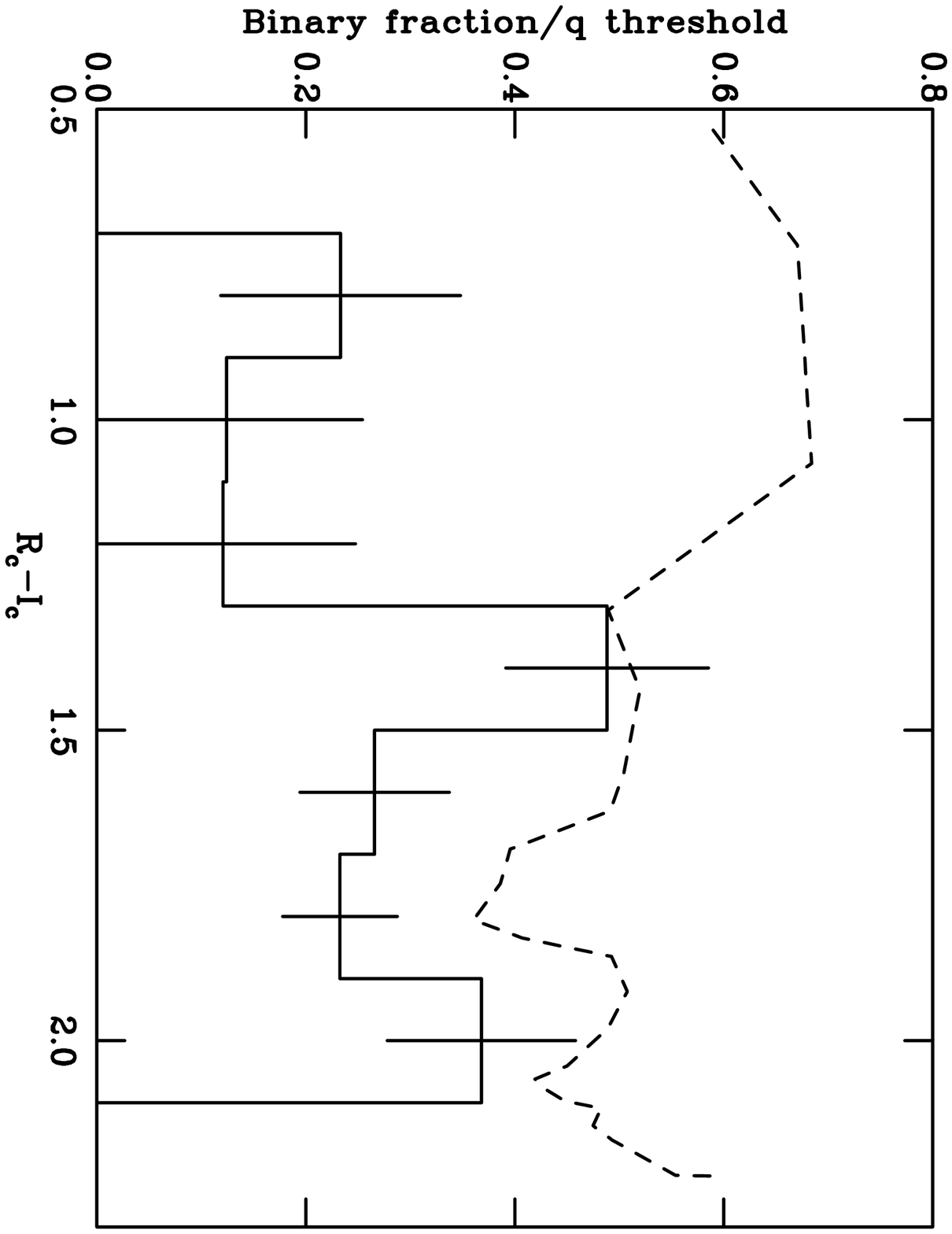}
\includegraphics{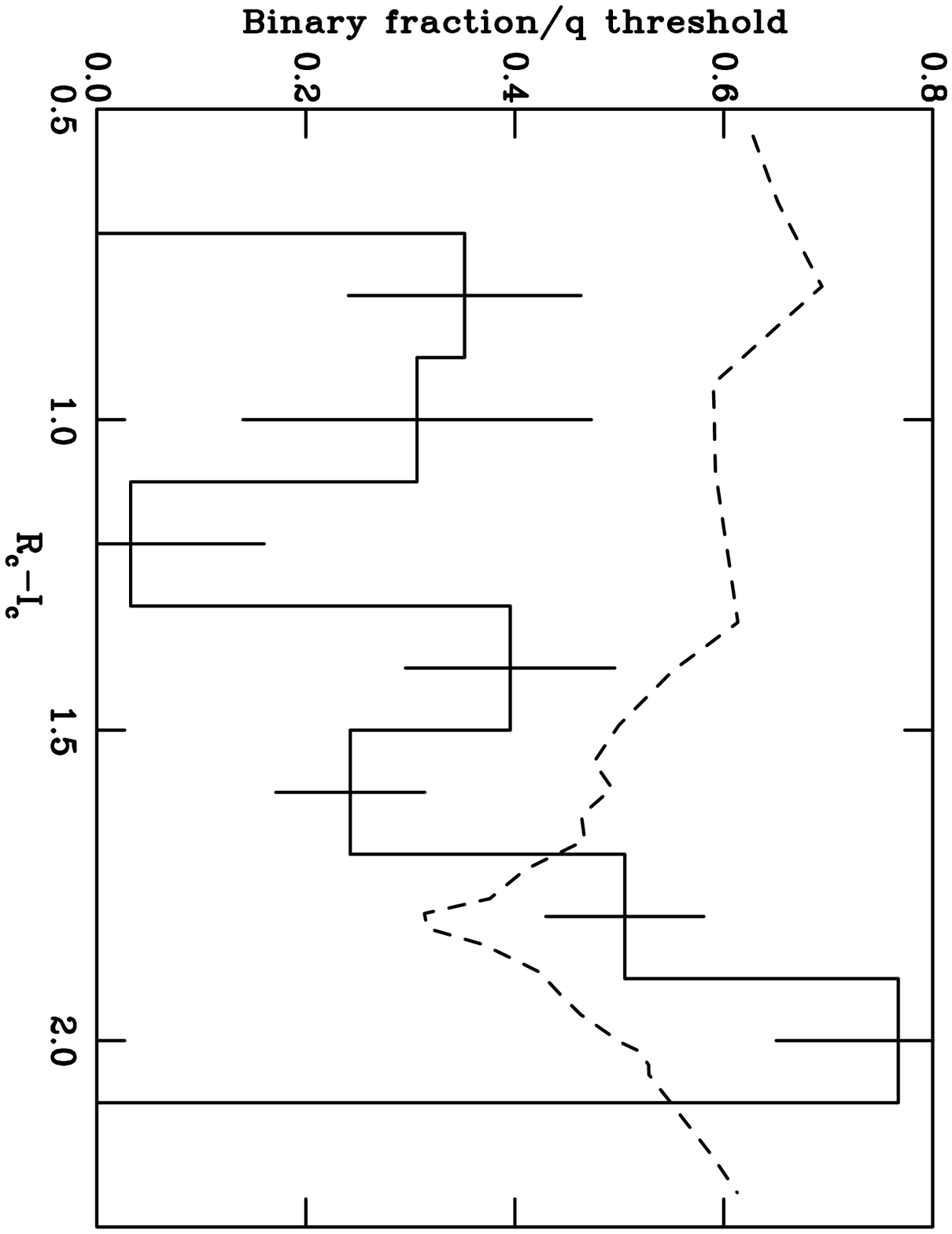}
\caption{(Top) The frequency of photometrically selected binary systems
(with $I_{\rm c}$ more than 0.5 mag above the single star isochrone) as
  a function of colour for the DAM97 isochrone selection (solid
  histogram with errors). The frequency
  is expressed as number of binary systems divided by total number of
  systems and has been corrected for completeness and contamination by
  field dwarfs. The threshold mass ratio ($q$) value corresponding to a
  0.5 mag $I_{\rm c}$ discrepancy from the single star sequence
  is shown as a dashed line. (Bottom) A
  similar plot when selection is made using the B02 isochrone.}
\label{binfrac}
\end{figure}

The binary frequency and distribution of binary periods and mass ratios
are important diagnostics of the star formation process (e.g. see
Halbwachs et al. 2003; Fisher 2004). 
In Figure~\ref{binfrac} we show the fraction of
stellar systems as a function of $R_{\rm c}-I_{\rm c}$ that are
identified as binary systems according to the photometric selection
techniques in described in section~\ref{select}. 
The binary fraction is defined as
the number of binary systems divided by the total number of
systems. The numbers of systems in each case were taken from within
a radius of 25 arcminutes of the cluster centre, corrected for completeness,
and a completeness-corrected number of contaminating field dwarfs from
a similar area was subtracted before performing the division.

The bluest bins are based on quite small number statistics and the very
bluest bin may contain a few contaminating giant stars. We also cannot
trust the results for $R_{\rm c}-I_{\rm c}\geq 2.0$ for several reasons: (i)
These stars will have $I_{\rm c}\geq18.5$ and the uncertainties in
their colours will confuse the sensitivty of our test for
binarity and bias it in a way that is difficult to predict (see
section~\ref{complete}) . (ii) At
these faint magnitudes there is a significant difference in the
completeness correction for binaries and single stars at the same
colour -- to the extent that many single stars would simply not have
been detected. The binary fraction is then heavily reliant on the
accuracy of the completeness corrections at low levels of completeness.
(iii) Our estimate of the sample contamination is likely to be too low
at these colours. This will likely {\it lower} the true binary fraction
because the fall off for the background density as a function of
decreasing magnitude at a given colour in the
CMD is quite steep.

For $0.9<R_{\rm c}-I_{\rm c}<1.9$ we obtain weighted mean binary
fractions of $0.25\pm0.07$ and $0.30\pm0.08$ when using the DAM97 and
B02 models respectively. There is marginal evidence that this fraction
is higher for $R_{\rm c}-I_{\rm c}>1.3$ (corresponding to masses of
approximately $<0.56\,M_{\odot}$ and $<0.58\,M_{\odot}$ according to
the DAM97 and B02 cluster isochrones we have used). This is
counterbalanced by the increasing range of $q$ values to which our
binary identification technique is sensitive (see Fig.~\ref{binfrac}),
which ranges (with some slight model dependency) from about
$0.65<q<1.0$ at $R_{\rm c}-I_{\rm c}\simeq 1.1$ to $0.35<q<1.0$ for
$R_{\rm c}-I_{\rm c}\simeq 1.9$, due to the changing slope of the
cluster single stars locus and the $I_{\rm c}$-mass relationship.

The deduced binary fraction for M dwarfs in NGC 2547 is similar
to binary fractions deduced for field M-dwarfs. Fischer \& Marcy (1992)
find that $42\pm9$ per cent of M-dwarf primaries form multiple systems
with $q>0.2$. Henry \& McCarthy (1990) find 34 per cent binarity for
M-dwarfs within 5.2\,pc, and the 8\,pc sample of M-dwarfs has 33 per
cent binarity (Reid \& Hawley 2000).  These latter two surveys almost
certainly suffer from a little incompleteness at the smallest mass
ratios. Comparable photometrically selected binarity statistics for
young clusters suggests a binary fraction of 26 per cent among the K
and early M stars of the Pleiades (with $q>0.4$ -- Stauffer et
al. 1984) and $36\pm5$ per cent for M5-M6 dwarfs in the Pleiades (with
$q>0.35$ -- Pinfield et al. 2003). We conclude that the binary fraction
for M-dwarfs in NGC 2547 is consistent with that in either the
field or the Pleiades.

\section{Discussion}
\label{discussion}

Most measurements of the initial mass function (IMF) to date have been
made in either very young star forming regions (e.g. IC\,348, Trapezium,
$\sigma$ Ori), where low-mass stars and brown dwarfs are intrinsically
luminous, or in older, rich clusters with large populations of low-mass
objects (e.g. Alpha Per, Pleiades, M\,35, NGC 2516). There are
disadvantages to both these methods. In the star forming regions one
must contend with the presence of discs, 
uncertainties in reddening and age as well as spreads
in these quantities. In addition there are likely systematic
uncertainties in low-mass evolutionary models due to to the
non-negligible influence of the initial conditions at very young ages
(B02). In older clusters ($>10$\,Myr) these particular systematic uncertainties
will be less, but there are then problems with mass segregation and the
evaporation of low-mass objects, uncertainties in the colours,
bolometric corrections and hence the magnitude-mass relationship for
very cool low-mass objects and the observational bias that
older clusters necessarily started off very rich and populous in order
to exist now in their current form. These very rich concentrations of
stars {\em may} have had a different IMF to that in less dense
star forming regions.

NGC 2547 forms an important bridge between these two categories of
objects. It is old enough that age spreads and reddening are not an
issue and systematic uncertainties in the evolutionary models are not
too important. On the other hand it is not old enough to have undergone
very significant mass segregation among its low-mass population and is
a smaller cluster than the Pleiades (see below).

\subsection{The IMF and NGC 2547}

The major result of this paper is that the MF of NGC 2547 is remarkably
consistent with that of the Pleiades down to the Hydrogen burning
limit (HBL). Just below the HBL there seems to be a dip in the MF of
NGC 2547 that is not present in the Pleiades, or in a number of other
young clusters. This dip is present at quite a high level of
significance -- for example, whereas we might have 
expected to detect about 10 objects in our NGC 2547 MF (using the DAM97
isochrones and multiplying by the likely level of incompleteness) 
with $-1.1 < \log M < -1.2$ if it had a similar MF to the
Pleiades, we only detected 2 such objects in NGC 2547,
both of which have a reasonably high probability of being contaminating
objects. Spreading the range to $-1.1< \log M < -1.3$ the comparison
becomes about 15 expected from a Pleiades MF versus 6 seen in 
NGC 2547 (of which about 
3 are expected to be contamination). At lower masses the MF of NGC 2547
appears to recover to levels seen in the Pleiades, but we cannot be
certain that this is the case, because our estimates of the contamination
level at colours corresponding to this mass in NGC 2547, are likely to
be too low (see section~\ref{contaminate}).

Our MF for NGC 2547 (and also the Pleiades) appears significantly lower (for
$M<0.1\,M_{\odot}$), than the log-normal form hypothesised 
to be universal among young disc clusters and star forming regions 
by Chabrier (2003). Because of its age NGC 2547 provides a unique
constraint on the form of the initial mass function for disc stars, but
we must consider several possible sources of systematic error.

\subsubsection{Methodology problems}

\begin{figure}
\vspace{70mm}
\includegraphics{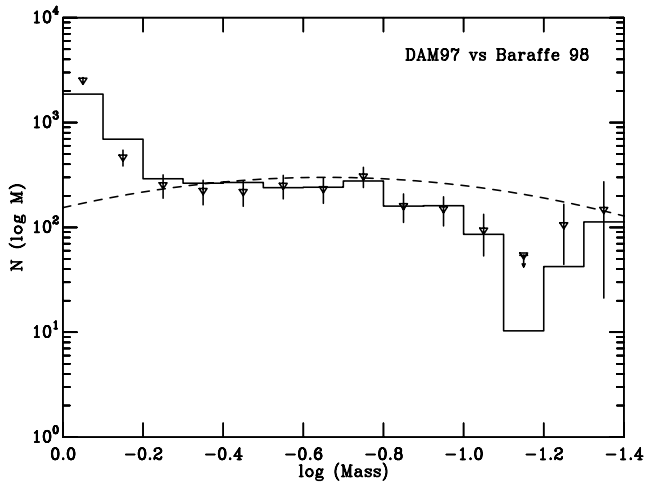}
\caption{The contamination-subtracted and completeness corrected MFs for
  NGC 2547 calculated using DAM97 isochrones at 30\,Myr (solid
  histogram) and a 32\,Myr $I_{\rm c}$ versus mass relation direct from
  the models of Baraffe et al. (1998 -- triangles with error bars).
  The dashed line shows the arbitrarily normalised young disc IMF
  proposed by Chabrier (2003).
}
\label{baraffemf}
\end{figure}

The possible problems that could have arisen in our analysis procedures
are centered upon (i) subtraction of background contamination; (ii)
correction for incompleteness and (iii) model dependence of the results.
\begin{enumerate}
\item A weakness in our analysis is that we were forced to rely upon a
  model to estimate the background contamination because we did
  not observe sufficient area at large distances from the cluster to
  make a precise
  observational estimate. However,
  the observations we have were 
  sufficient to show that our background estimates
  are not likely to be wrong by factors of two, except perhaps at the
  very lowest masses $\leq 0.05\,M_{\odot}$, where we make no claim to have
  accurately determined the NGC 2547 MF. 
  Because of the relatively high contrast
  between the cluster and background at $r<25$ arcminutes,
  uncertainties in the contamination cannot change our main conclusions,
  though they could alter the exact slope of the MF at low masses,
  particularly if the B02 isochrones (for which the low-mass cluster
  members are more heavily contaminated by background) turn out to be
  more appropriate than those of DAM97.
  If anything, we believe our model {\em
  underpredicts} the amount of contamination at the lowest masses,
  which would lead to fewer brown dwarfs in NGC 2547.

\item We have made great efforts to ensure our completeness corrections
  are as accurate as possible -- including a number of effects rarely
  addressed in the literature to date (for example the spatial and
  colour dependence of completeness and the effects of non-Gaussian
  uncertainties which scatter stars sufficiently for them to be excluded as
  candidate cluster members). Our survey is substantially complete
  ($>60$ per cent for $I_{\rm c}<19.5$, corresponding to $M\simeq
  0.06\,M_{\odot}$. To explain the discrepancy between NGC 2547 and the
  log-normal disc MF of Chabrier (2003) 
  would require these completeness corrections to be {\em smaller}
  by {\em factors} of 2-3 for $M\leq 0.1\,M_{\odot}$ ($I_{\rm
  c}\geq18.4$). Systematic errors of this size seem most unlikely (see
  for instance the tests performed in N02).

\item We have used those readily available models which cover the
  required mass range in our data. There are some model dependencies in
  the results -- notably that the DAM97 models do allow the possibility
  of some mass segregation at low-masses, while the B02 models do not.
  These differences are as a result of the slightly differing candidate
  members selected using the two models and the concomitant differences
  in the estimated levels of contamination.
  However, the magnitude-mass relationships are remarkably similar and
  so the MF of NGC 2547 is essentially the same for both sets of
  models. These magnitude-mass relationships make use of empirical
  relationships between colour and bolometric correction and an
  assumed age and distance for the Pleiades. An alternate approach
  adopted by many authors is to use the theoretical 
  bolometric corrections and magnitudes directly calculated in the
  Baraffe et al. (1998) {\sc nextgen} models to estimate masses.
  To test for sytematic error, we recalculated the MF of NGC 2547 using
  this technique. The result is shown in
  Fig.~\ref{baraffemf} and illustrates very little sensitivity to
  this change. This is probably because there is very little
  disagreement between the theoretical bolometric corrections of
  Baraffe et al. (1998) and the empirical bolometric corrections
  adopted in this paper over the colour range of interest for NGC 2547.
  This is not likely to be the case for brown dwarfs in older clusters,
  where photospheric temperatures may be cool enough for dust
  formation and hence bluer colours and fainter magnitudes for the same mass.
  However, Dobbie et al. (2002) have suggested this may be a problem even at
  temperatures as warm as 2700\,K (corresponding to $R_{\rm c}-I_{\rm
  c}\simeq 2.15$ or 2.30 for the B02 or DAM97 model isochrones we have
  used), causing a dip in MFs derived from models that do not
  incorporate these effects (see section~5.3). 
  Such temperatures are present at the faint
  end of our NGC 2547 sample and we cannot be sure that our MFs are not
  susceptible to sytematic effects ignored (or included incompletely)
  in all available models.
\end{enumerate}

\subsubsection{Hidden binary companions}

In section~\ref{mf}, we pointed out the problem of unresolved binarity
in determining the true MF. We ignored the problem, with the caveat
that, when making comparisons with the field and other clusters, we
must assume that the binary frequency and $q$ distribution (as a
function of mass) is similar. A recent series of papers (Kroupa \&
Bouvier 2003; Kroupa et al. 2003) have highlighted that differences in
binary properties can lead to {\em apparent} differences in the
observed MF. The particular example discussed is the apparent deficit
of brown dwarfs in the Taurus-Auriga association (TA) compared with the
Orion Nebula Cluster (ONC -- see Brice\~{n}o et al. 2002). 
The hypothesis is that many
``missing'' brown dwarfs are still locked in binary systems in the TA,
while these binaries have been disrupted in the denser environs of the ONC.

So a possible solution to the apparent deficit of high-mass brown
dwarfs in NGC 2547 compared to the Pleiades could be an enhanced binary
fraction, especially of low-$q$ systems. The results discussed in
section~\ref{binary} partly argue against this possibility. The binary
frequencies of $q>0.4$ binaries with primary masses $0.1<M<0.3$ are
similar to those in the Pleiades.  However, there could still be an
anomalously high number of $q<0.4$ systems in NGC 2547, which we could
not identify with our photometric techniques and in any case we have
not been able to estimate a binary fraction below the HBL.

\subsubsection{Incorrect cluster age}

\begin{figure}
\vspace{70mm}
\includegraphics{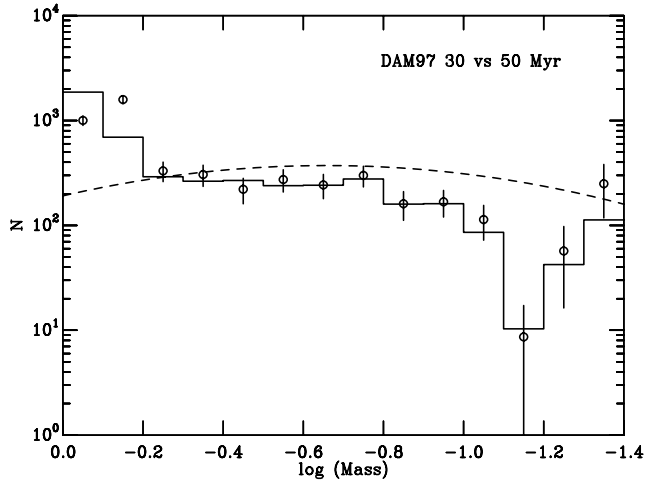}
\caption{The contamination-subtracted and completeness corrected MFs for
  NGC 2547 calculated using DAM97 isochrones at 30\,Myr and a distance
  modulus of 8.1 (solid line) compared with 50\,Myr and a distance
  modulus of 7.7 (points with error bars).
}
\label{30vs50}
\end{figure}

The low-mass isochronal age of NGC 2547 is about 30\,Myr, but
there has been some debate in the literature concerning this age. 
Clari\'a (1982) and Jeffries \& Tolley (1998) find an age of
about 50\,Myr from the high-mass turn-off in the cluster, albeit
defined by only one or two stars. Jeffries et al. (2003) and Oliveira
et al. (2003) conclude that the pattern of lithium depletion seen amongst
the low-mass members of NGC 2547 may also be best explained (using the {\it
  same} models used to fit isochronal ages)  if the cluster age is
$50$\,Myr. A similar pattern is exhibited in other young clusters,
where the isochronal age (both low and high mass) is found to be
younger than the age determined from Li depletion mechanisms
(e.g. Barrado y Navascu\'{e}s, Stauffer \& Patten 1999).

Our low-mass data are incapable of simultaneously fixing the age and
distance of NGC 2547, because at ages around 30\,Myr the isochrones are
almost parallel (certainly for $R_{\rm c}-I_{\rm c}>0.5$ 
see Figs.~\ref{xrayricmd1} and~\ref{xrayricmd2}). 
Hence a change of age can be compensated by a
change in distance modulus (considering only the low-mass data
presented here). To check the age dependence of our MFs we increase
the assumed age of the cluster to 50\,Myr and {\it decrease} the distance
modulus to 7.8 and 7.7 in the case of the B02 and DAM97 models
respectively. These isochrones yield new $I_{\rm c}$-mass relationships
that can then be used to generate MFs.

We find that after correction for the differing distance moduli, the
mass-magnitude relationships are almost identical over the mass range
of interest here. Figure~\ref{30vs50} shows a comparison of the MFs
derived assuming a cluster age of 30\,Myr and 50\,Myr and the DAM97
evolutionary models. Uncertainty in the cluster age at this
level cannot affect any of our conclusions.

\subsubsection{Mass segregation}

If mass segregation is present in the cluster then a survey that does
not extend to the cluster tidal radius may underestimate the
contribution from low-mass stars/brown dwarfs. An example of this
is M\,35, a cluster which is old enough (175\,Myr) to be dynamically
relaxed (see below), shows a sharp downturn in its MF below
0.2\,$M_{\odot}$ from a survey of limited area (Barrado y Navascu\'{e}s
et al. 2001), and which shows clear evidence for mass segregation
(Kalirai et al. 2003).

The analysis in section~\ref{masseg} failed to reveal significant
mass segregation in NGC 2547 between samples with mean masses of
0.16\,$M_{\odot}$ and 0.50\,$M_{\odot}$. Mass segregation of the 
form $r_{c} \propto M^{-0.5}$ is expected in the case of energy
equipartion (Pinfield et al. 1998). We can rule this out in the case where
we use the B02 isochrones and it very unlikely when using the DAM97
isochrones. 

An approximate calculation can be used to check whether this
is reasonable. Jeffries et al. (2000) measured a 1-dimensional
velocity dispersion of $<0.9$\,km\,s$^{-1}$ for solar-mass stars in
NGC~2547, although given that this
is similar to the esimated radial velocity measurement errors, this
must be a very conservative upper limit. This gives a crossing time-scale
(for 2 core radii) of $>4\,$Myr. The number of stars/systems
($>0.1\,M_{\odot}$) in the cluster is $\sim 500$ and using Binney \&
Tremaine's (1987) approximate formula
\be
t_{\rm relax} \simeq t_{\rm cross}\, \left(\frac{N}{8 \ln N}\right)\, ,
\ee
an estimate for the relaxation timescale is $>40$\,Myr. 

Numerical simulations for general clusters and specifically for NGC
2547 confirm these approximate results (e.g. Adams et al. 2002,
Littlefair et al. 2003).  
At an age of about 30\,Myr,
little mass segregation is expected in NGC 2547. That which is, should
affect the highest mass stars first. Hence any major mass segregation
seen, such as that obvious between stars with $M>3\,M_{\odot}$ and the
rest of the cluster, is most likely primordial in nature.  
The relatively constant behaviour of the cluster core
radius for stars from $0.1<M<0.7 M_{\odot}$) {\em does not} support any
suggestion that they were formed with a mass-dependent spatial
distribution or velocity dispersion.

What little evidence for mass segregation that exists (for the DAM97
models) would have very little effect on the measured MF. Extrapolating
the King profiles to the tidal radius we see that the ratio of
0.16\,$M_{\odot}$ to 0.50\,$M_{\odot}$ stars will only increase by a
factor of $\simeq 1.05$ from that measured within a radius of 25
arcminutes.

\subsubsection{Evaporation of low-mass objects}

N-body simulations show that the process of dynamical relaxation leads
to the preferential evaporation of low-mass objects, in addition to
mass segregation (e.g. de la Fuente 2000). 
Adams et al. (2002) show that {\it dynamical} evaporation is unlikely to be
a factor until cluster ages significantly greater than the 
relaxation time. As there is little sign of mass segregation we can 
safely conclude that dynamical evaporation is also unlikely to be
important in NGC 2547.

However, several authors have put forward models where very low-mass
stars and especially brown dwarfs are preferentially expelled from
protostellar multiple systems, before they have had chance to accrete
significant masses - the ``ejection hypothesis'' -- (e.g. Reipurth \&
Clarke 2001; Sterzik \& Durisen 2003). As a result they are likely to
have a larger initial velocity dispersion than higher mass stars,
resulting in preferential mass segregation and evaporation from a
cluster.  A number of problems have emerged with the ``ejection
hypothesis''. For example, the spatial distribution of low-mass stars
and brown dwarfs are found to be similar in some young star forming
regions (e.g. L\'{o}pez-Mart\'i et al. 2004).  Our results cannot
directly contradict the ejection hypothesis, because there are too few
objects with $M<0.1\,M_{\odot}$ to assess their spatial distribution.
The escape velocities for low-mass objects in NGC 2547 and the Pleiades
(using the cluster masses derived below) are quite similar at $\sim
1$\kms, so a similar fraction of brown dwarfs might be expected to have
escaped in this way.

The early ejection of brown dwarfs from a cluster would result in a
cluster MF significantly depleted of brown dwarfs compared with the
field. Unfortunately, constraints on the field brown dwarf MF are still
too loose to pinpoint any inconsistency between cluster and field MFs
at the lowest masses (e.g. Chabrier 2003).

\subsection{The mass of NGC 2547}
\label{clustermass}

The total mass of NGC 2547 can be estimated by summing the number of
stellar systems estimated within the tidal radius ($n_{t}$ in
Table~\ref{kingtable}) multiplied by their average mass, which comes to
about 350\,$M_{\odot}$. We need to add an extra contribution due to NGC
2547 stars that lie within the magnitude range heavily contaminated by
giants. Assuming that the MF is roughly flat and has
$dN/d \log M \simeq 300$ between $\log M=-0.15$ (corresponding to $I_{\rm
c}\simeq 14.5$) and $\log M=+0.05$ (corresponding to $V\simeq13.0$),
multiplying by a factor of 1.25 to approximately account for binarity
(50 per cent binarity and a flat $q$ distribution) and a factor of 1.6
to account for stars out to the tidal radius, we arrive at an extra
100\,$M_{\odot}$ in this mass range. The total cluster is therefore
about 450$\,M_{\odot}$ for $M>0.057\,M_{\odot}$\footnote{Which is what
we assumed in order to calculate the tidal radius in
section~\ref{masseg}. In practice, as the reader might have guessed, it
required an iteration of the analysis to arrive at this consistency.},
which can be compared with a total Pleiades mass of 735\,$M_{\odot}$
derived by Pinfield et al. (1998).
Uncertainties in the shape of the mass function, choice of evolutionary
isochrones and uncertainties in the King profile corrections make the
total mass of NGC 2547 uncertain by no more than $\pm 100\,M_{\odot}$.

\subsection{Possible Variations in the IMF?}

Although the MF of NGC 2547 is almost identical to the Pleiades down to
the HBL, we have tentatively identified a deficit of high-mass brown
dwarfs in NGC 2547. We can go further and state that, because our
contamination model suffers an unknown level of incompleteness for
masses less than about $0.06\,M_{\odot}$ in NGC 2547, our analysis
(using either the DAM97 or B02 models) is consistent with there being
{\em no} brown dwarfs in NGC 2547.  Clearly this requires further
testing. In particular, a deeper, wider survey, perhaps incorporating
near-IR data, would provide a much more secure empirical estimate of
the amount of background contamination at very red colours.

Many believe that the IMF in young clusters and associations may be
universal (e.g. Kroupa 2002), but evidence is now emerging for
significant differences in the measured MFs in differing star forming
regions. The Taurus-Auriga and IC\,348 young star forming regions both
appear to show a factor of two or more deficit in the ratio of brown
dwarfs to low-mass stars, when compared with regions like the Pleiades
or the ONC (Brice\~{n}o et al. 2002; Luhman et al. 2003, Preibisch,
Stanke \& Zinnecker 2003). In particular, the MF for IC\,348 presented
by Luhman et al. (2003) shows a sharp 0.6\,dex fall below
$0.1\,M_{\odot}$ that is quite reminiscent of what we have found for
NGC 2547.  These authors speculate that either the low stellar density
in these regions or the absence of very high mass stars
($>6\,M_{\odot}$) may be important factors. It could be that the
far-ultraviolet radiation from hot stars is an important factor in
terminating accretion in low-mass objects, resulting in more brown
dwarfs.

The lack of brown dwarfs in NGC 2547 could be interesting in this
context. The most massive star in the cluster has $M<6\,M_{\odot}$ and
an extrapolation of the high-mass end of the present-day MF suggests
that more massive stars may never have been present (see N02). 

An alternative explanation might involve an incomplete understanding of
cool atmospheres in low-mass stars. Dobbie et al. (2002) have
identified a number of ``dips'' in the LFs of young clusters (Pleiades,
$\sigma$ Ori, Alpha Per, IC\,2391) which, whilst individually not very
significant, collectively seem to point to a drop in the LF (and
consequently the MF) at effective temperatures of about 2700\,K.
Dobbie et al. interpret this as the onset of dust formation with grains
somewhat larger than assumed in most models, which would lead to
enhanced opacity at near infrared wavelengths and a modified
mass-magnitude relationship. Our observations may support this
idea. The ``gap'' in the CMD seen in Fig.~7 occurs at colours
corresponding to 2800-2900\,K in the B02 and DAM97 models. Given the
likely systematic errors in $R_{\rm c}-I_{\rm c}$ this could be consistent
with the 2700\,K value proposed by Dobbie et al. If the LF and MF of
NGC 2547 really did increase again for $I_{\rm c}>19.5$ then NGC 2547
might be the clearest example yet of this phenomenon.

\section{Conclusions}

In this paper we have presented optical/near infrared photometry of NGC
2547 that is both deeper and covers a wider area (0.855 square degrees)
than previously published surveys. The survey has been used to collate
a catalogue of candidate members and investigate the degree of mass
segregation and the mass function (MF) for the cluster.  The main
conclusions of our work can be summarised as follows:

\begin{enumerate}

\item The cluster shows strong evidence for mass segregation, in that
  stars with $M>3\,M_{\odot}$ are {\it much} more centrally
  concentrated than lower mass stars. There is some evidence for mass
  segregation above and below 1\,$M_{\odot}$, but lower mass stars have
  spatial distributions that are consistent with no further mass
  segregation for $0.1<M<0.7\,M_{\odot}$. By fitting King profiles we
  have estimated that at least 60 per cent of the cluster's
  low-mass population is included within our survey.

\item The MF for $0.075<M<0.7\,M_{\odot}$ in NGC 2547 has been
  determined with roughly the same statistical precision as available
  for the best Pleiades studies (and much more precisely than field MF
  determinations) and is in remarkable agreement with the Pleiades MF
  (and other young clusters) over this mass range.  Because of it's age
  ($\simeq 30$\,Myr), NGC 2547 is not badly affected by a number of
  systematic uncertainties which could well affect studies of both
  younger and older clusters (age spreads and uncertainties, reddening,
  discs, very cool
  atmospheres, dynamical evolution) and neither are our results very
  dependent on which stellar evolutionary models are used. Our analysis
  therefore represents one of the most precise and robust
  determinations of a low-mass disc population MF, and possibly the initial MF
  if primordial velocity dispersions are reasonably mass-independent.

\item Below the hydrogen burning limit we have identified an apparent dearth of
  high-mass brown dwarf candidates in NGC 2547
  ($0.05<M<0.075\,M_{\odot}$). This deficit {\it may} extend to lower
  masses, but increasing incompleteness in both our survey and our
  estimates of contaminating foreground M-dwarfs leave this question
  open. Alternatively, we may be seeing an example of the ``missing M
  dwarf'' phenomenon, identified by Dobbie et al. (2002), which could
  result from an imperfect understanding of atmospheres with
  temperatures of around 2700\,K and would not reflect a true deficit of
  brown dwarfs.  A deeper, wider survey of NGC 2547 is
  called for to verify this result, to provide a better empirical estimate
  of contamination among the lowest mass cluster candidates, and hence
  show whether the deficit we have seen is merely a dip or a
  genuine lack of brown dwarfs.

\item The total mass of NGC 2547 for stars with $M>0.06\,M_{\odot}$ is
  $(450\pm 100)\,M_{\odot}$ and about a factor of two smaller than the Pleiades.

\item The binary fraction of M-dwarfs in NGC 2547 is between 20 and 35
  per cent for systems with mass ratios greater than 0.35 to
  0.65. This fraction is consistent with values determined for
  populations of low-mass stars found in the field and other young
  clusters.

\item Finally, we have provided in electronic format both our entire
 photometric catalogue as well as subsets of photometrically selected
 cluster candidates.  These catalogues contain data with robust and
 precisely determined photometric and astrometric uncertainties.
 
\end{enumerate}

\section*{Acknowledgments}
We would like to thank the director and staff of the Cerro Tololo
Interamerican Observatory, operated
by the Association of Universities for Research in Astronomy, Inc.,
under contract to the US National Science Foundation.
Computing was performed
at the Keele and Exeter nodes of the Starlink network, funded by PPARC.
EJT was supported during this work from a PPARC research grant.
CRD was supported by a Nuffield Undergraduate Bursary (NUF-URB00).
The defringing algorithm used in our data reduction was kindly supplied
by Mike Irwin. We thank the Paul Dobbie for supplying a very prompt and
useful report.

The Digitized Sky Survey was produced at the Space Telescope Science
Institute under U.S. Government grant NAG W-2166. The images of these
surveys are based on photographic data obtained using the Oschin
Schmidt Telescope on Palomar Mountain and the UK Schmidt Telescope. The
plates were processed into the present compressed digital form with the
permission of these institutions.

\nocite{dobbie02}
\nocite{binney87}
\nocite{kroupa02}
\nocite{adams02}
\nocite{barrado99}
\nocite{barrado01}
\nocite{stauffer84}
\nocite{reidhawley00}
\nocite{sagar91}
\nocite{lopez04}
\nocite{kalirai03}
\nocite{fisher04}
\nocite{jeffries03}
\nocite{oliveira03}
\nocite{fischer92}
\nocite{kroupabouvier03}
\nocite{kroupa03}
\nocite{preibisch03}
\nocite{briceno02}
\nocite{luhman03}
\nocite{henry90}
\nocite{jameson02}
\nocite{barrado02}
\nocite{bejar01}
\nocite{reid02}
\nocite{halbwachs03}
\nocite{sagar91}
\nocite{bouvier01}
\nocite{pinfield98}
\nocite{king62}
\nocite{chabrier03}
\nocite{sterzik03}
\nocite{reipurth01}
\nocite{naylor02}
\nocite{bonnell01}
\nocite{claria82}
\nocite{jeffries98n2547}
\nocite{littlefair03}
\nocite{jeffries00}
\nocite{jth01}
\nocite{zapatero99}
\nocite{oliveira03}
\nocite{baraffe02}
\nocite{baraffe98}
\nocite{dantona97}
\nocite{leggett92}
\nocite{leggett96}
\nocite{stauffer82}
\nocite{bouvier98}
\nocite{moraux01}
\nocite{moraux03}
\nocite{hambly01a}
\nocite{hambly01b}
\nocite{bate03}
\nocite{burningham03}
\nocite{bessell87}
\nocite{branham03}
\nocite{gould97}
\nocite{meusinger96}
\nocite{pinfield03}
\nocite{delafuente00}
\nocite{schaller92}
\nocite{kroupa01}

\bibliographystyle{mn2e}  
\bibliography{iau_journals,master}

\appendix
\section{Catalogues}

Three electronic catalogues are available from this paper. The first is
our entire photometric/astrometric catalogue, containing 133788
entries. A sample is shown below in Table~A1 as a guide to
its form and content. We also provide subsets of this catalogue which
contain candidate members of NGC 2547, selected according to the
criteria in section~\ref{select}.  Table~A2 lists thise
candidate members selected using the B02 isochrone, while
Table~A3 lists those selected using the DAM97
isochrone. The two catalogues contain 726 and 744 entries respectively,
with 682 objects common to both. Note that these catalogues suffer from
incompleteness and contamination as described in
sections~\ref{complete} and~\ref{contaminate}.

The format of these catalogues is one row per star consisting of a
field number (note though that a star may have been detected in more
than one field because of the overlaps), star identifier in that field,
x and y pixel positions on a reference frame in that field, right
ascension and declination (J2000.0), and three sets of magnitude,
magnitude error and magnitude flags corresponding to the $I_{\rm c}$,
$R_{\rm c}-I_{\rm c}$ and $I_{\rm c}-Z$ values respectively. The flags
consist of two alphabetic characters for each magnitude, referring to
the first and second measurements that contribute to the final
magnitude. Objects with no problems in their detection or measurement
are flagged ``OO'', the other flags are described in Burningham et
al. (2003).

The catalogues can be obtained from the online version of the journal
on {\it Synergy}, from the Centre de Donn\'{e}es astronomiques de Strasbourg or the
Cluster Collaboration web page at\\
www.astro.ex.ac.uk/people/timn/Catalogues/description.html

\newpage
\label{lastpage}
\begin{landscape}
\textwidth=240mm
\begin{table*}
\label{catalogue}
\caption{A sample of the full photometric and astrometric catalogue for NGC 2547. The
  full table is only available electronically.}
  \begin{tabular}{@{}cccc..cccccccccc@{}}
                   &      & 
\multicolumn{2}{c}{J2000 Position} & 
\multicolumn{2}{c}{Pixel Position} & 
\multicolumn{3}{c}{$I_{\rm c}$} &
\multicolumn{3}{c}{$R_{\rm c}-I_{\rm c}$} &
\multicolumn{3}{c}{$I_{\rm c}-Z$} \\
 Field & Star &                  
R.A. & Dec. &                
X & Y &
Mag. & Error & Flag  & 
Mag. & Error & Flag  & 
Mag. & Error & Flag  \\ 
  51 &   3612 & 08 04  5.376& -48 24 16.58&  2036.691&    66.065 &    25.449 &     0.755 & OE &  -5.519 &     0.735&  OE &  -0.167 &     0.904&  EE\\
  51 &   4639 & 08 04  5.386& -48 28 11.35&  2034.791&   651.172 &    19.949 &     0.105 & OO &   0.798 &     0.113&  OO &  -0.180 &     0.196&  OO\\
  51 &   5500 & 08 04  5.386& -48 25 56.23&  2035.732&   314.446 &    20.910 &     0.250 & OO &   0.574 &     0.261&  OO &   0.789 &     0.311&  OO\\
  51 &   3903 & 08 04  5.396& -48 29 44.64&  2033.890&   883.673 &    19.030 &     0.044 & OO &   0.421 &     0.046&  OO &   0.108 &     0.081&  OO\\
  51 &   2055 & 08 04  5.422& -48 30 13.26&  2033.065&   955.029 &    17.524 &     0.021 & OO &   0.540 &     0.023&  OO &   0.273 &     0.033&  OI\\
  51 &    914 & 08 04  5.445& -48 24 58.94&  2034.706&   171.642 &    17.425 &     0.026 & NN &   0.413 &     0.029&  NN &   0.161 &     0.033&  NN\\
  51 &   4781 & 08 04  5.484& -48 30 27.57&  2031.394&   990.664 &    19.693 &     0.098 & OO &   0.061 &     0.110&  OO &   0.614 &     0.142&  OI\\
  51 &   3497 & 08 04  5.504& -48 35 53.37&  2028.676&  1802.636 &    19.215 &     0.052 & OO &   0.745 &     0.056&  OO &   0.338 &     0.069&  OO\\
  51 &   4520 & 08 04  5.527& -48 26 23.64&  2032.090&   382.744 &    20.322 &     0.126 & OO &   0.869 &     0.141&  OO &   0.026 &     0.211&  OO\\
  51 &   1225 & 08 04  5.553& -48 27 18.73&  2031.014&   520.017 &    17.736 &     0.015 & OO &   0.394 &     0.016&  OO &   0.138 &     0.023&  OO\\
  51 &   3702 & 08 04  5.589& -48 26  1.73&  2030.705&   328.098 &    18.685 &     0.067 & NN &   0.765 &     0.075&  NN &   0.276 &     0.082&  NN\\
  51 &   1446 & 08 04  5.602& -48 36 32.15&  2025.960&  1899.284 &    17.595 &     0.032 & NN &   0.440 &     0.034&  NN &   0.158 &     0.040&  NN\\
  51 &   1105 & 08 04  5.606& -48 35 39.95&  2026.239&  1769.169 &    17.266 &     0.011 & OO &   0.556 &     0.012&  OO &   0.157 &     0.016&  OO\\
  51 &   5222 & 08 04  5.609& -48 36 55.65&  2025.646&  1957.868 &    19.032 &     0.099 & NN &   4.309 &     0.755&  NN &  -2.119 &     0.154&  NN\\
  51 &   6203 & 08 04  5.622& -48 33  1.15&  2026.984&  1373.438 &    20.080 &     0.107 & OO &   1.018 &     0.124&  OO &   0.408 &     0.140&  OO\\
  51 &    552 & 08 04  5.629& -48 26 40.95&  2029.409&   425.844 &    16.429 &     0.006 & OO &   0.711 &     0.007&  OO &   0.212 &     0.009&  OO\\
  51 &   2619 & 08 04  5.632& -48 30 47.40&  2027.668&  1040.092 &    18.994 &     0.044 & OO &   0.460 &     0.046&  OO &   0.072 &     0.072&  OO\\
  51 &   1854 & 08 04  5.635& -48 24 25.17&  2030.183&    87.482 &    18.151 &     0.046 & NN &   0.549 &     0.050&  NN &   0.131 &     0.059&  NN\\
  51 &   2059 & 08 04  5.638& -48 30 21.57&  2027.665&   975.724 &    17.768 &     0.037 & IO &   0.497 &     0.040&  IO &   0.267 &     0.045&  OI\\
  51 &   2194 & 08 04  5.642& -48 34 20.88&  2025.907&  1572.124 &    18.413 &     0.057 & NN &   0.835 &     0.064&  NN &   0.580 &     0.060&  NN\\
  51 &    423 & 08 04  5.668& -48 26  3.84&  2028.733&   333.376 &    16.348 &     0.006 & OO &   0.554 &     0.007&  OO &   0.183 &     0.008&  OO\\
\end{tabular}
\end{table*}

\begin{table*}
\label{membar19}
\caption{A sample of the subset of the full catalogue containing candidate members of
  NGC 2547 selected using a Baraffe et al. (2002) isochrone (see
  section~\ref{select}). The full
  table is only available electronically.}
  \begin{tabular}{@{}cccc..cccccccccc@{}}
                 &      & 
\multicolumn{2}{c}{J2000 Position} & 
\multicolumn{2}{c}{Pixel Position} & 
\multicolumn{3}{c}{$I_{\rm c}$} &
\multicolumn{3}{c}{$R_{\rm c}-I_{\rm c}$} &
\multicolumn{3}{c}{$I_{\rm c}-Z$} \\
 Field & Star &                  
R.A. & Dec. &                
X & Y &
Mag. & Error & Flag  & 
Mag. & Error & Flag  & 
Mag. & Error & Flag  \\ 
  51&   48&    8  4 10.205&  -48 31 26.50&  1914.251&  1137.169 &    13.068 &     0.003&   OO &   0.570 &     0.004&   OO &   0.190 &     0.004 &  OO\\
  51&  260&    8  4 12.936&  -48 27 12.39&  1848.160&   503.670 &    15.352 &     0.004&   OO &   1.246 &     0.005&   OO &   0.326 &     0.005 &  OO\\
  51& 4399&    8  4 14.176&  -48 24 35.24&  1818.334&   111.949 &    20.033 &     0.087&   OO &   2.421 &     0.196&   OO &   0.682 &     0.100 &  OO\\
  51&   31&    8  4 17.313&  -48 28 23.15&  1739.295&   679.771 &    12.827 &     0.003&   OO &   0.363 &     0.007&   OO &   0.093 &     0.004 &  OO\\
  51& 2012&    8  4 19.765&  -48 28 58.81&  1678.399&   768.470 &    18.473 &     0.025&   OO &   1.786 &     0.039&   OO &   0.579 &     0.029 &  OO\\
  51&   55&    8  4 23.267&  -48 24 43.58&  1592.836&   132.166 &    13.799 &     0.003&   OO &   0.705 &     0.004&   OO &   0.250 &     0.004 &  OO\\
\end{tabular}
\end{table*}

\begin{table*}
\label{memdam97}
\caption{A sample of the subset of the full catalogue containing candidate members of
  NGC 2547 selected using a D'Antona \& Mazzitelli (1997) isochrone
  (see section~\ref{select}). The full
  table is only available electronically.}
  \begin{tabular}{@{}cccc..cccccccccc@{}}
       &      & 
\multicolumn{2}{c}{J2000 Position} & 
\multicolumn{2}{c}{Pixel Position} & 
\multicolumn{3}{c}{$I_{\rm c}$} &
\multicolumn{3}{c}{$R_{\rm c}-I_{\rm c}$} &
\multicolumn{3}{c}{$I_{\rm c}-Z$} \\
Field & Star &                  
R.A. & Dec. &                
X & Y &
Mag. & Error & Flag  & 
Mag. & Error & Flag  & 
Mag. & Error & Flag  \\ 
  51&   48 &   8  4 10.205&  -48 31 26.50&  1914.251 & 1137.169 &    13.068 &     0.003 &  OO &   0.570 &     0.004&   OO &   0.190 &     0.004 &  OO\\
  51&  260 &   8  4 12.936&  -48 27 12.39&  1848.160 &  503.670 &    15.352 &     0.004 &  OO &   1.246 &     0.005&   OO &   0.326 &     0.005 &  OO\\
  51& 4399 &   8  4 14.176&  -48 24 35.24&  1818.334 &  111.949 &    20.033 &     0.087 &  OO &   2.421 &     0.196&   OO &   0.682 &     0.100 &  OO\\
  51&  106 &   8  4 15.530&  -48 27 30.30&  1783.775 &  548.145 &    14.182 &     0.003 &  OO &   0.634 &     0.004&   OO &   0.212 &     0.004 &  OO\\
  51&   31 &   8  4 17.313&  -48 28 23.15&  1739.295 &  679.771 &    12.827 &     0.003 &  OO &   0.363 &     0.007&   OO &   0.093 &     0.004 &  OO\\
  51&   55 &   8  4 23.267&  -48 24 43.58&  1592.836 &  132.166 &    13.799 &     0.003 &  OO &   0.705 &     0.004&   OO &   0.250 &     0.004 &  OO\\
\end{tabular}
\end{table*}

\end{landscape}

\end{document}